\documentclass[aps,showpacs,floats,superscriptaddress,preprint]{revtex4}  
\usepackage{amsmath}
\usepackage{amssymb}
\usepackage{graphicx}
\def\simle{
    \mathrel{\rlap{\raise 0.511ex
        \hbox{$<$}}{\lower 0.511ex \hbox{$\sim$}}}}

\begin{document}

\title{
Study of the slepton non-universality at the CERN Large Hadron Collider
}

\author{Toru~Goto}
\email{goto@yukawa.kyoto-u.ac.jp}
\affiliation{YITP, Kyoto University, Kyoto 606-8502, Japan}

\author{Kiyotomo Kawagoe}
\email{kawagoe@phys.sci.kobe-u.ac.jp}
\affiliation{Department of Physics, Kobe University, Kobe 657-8501, Japan}

\author{Mihoko~M.~Nojiri}
\email{nojiri@yukawa.kyoto-u.ac.jp}
\affiliation{YITP, Kyoto University, Kyoto 606-8502, Japan}

\begin{abstract}
In supersymmetric theory, the sfermion-fermion-gaugino interactions
conserve the chirality of (s)fermions.
The effect appears as the charge asymmetry in $m(jl)$ distributions at the
CERN Large Hadron Collider where jets and leptons arise from the cascade
decay $\tilde{q} \to q \tilde{\chi}^0_2 \to ql\tilde{l}$.
Furthermore, the decay branching ratios and the  charge asymmetries
in $m(jl)$ distributions are flavor non-universal due to the
$\tilde{l}_L$ and $\tilde{l}_R$ mixing.
When $\tan\beta$ is large, the non-universality between $e$ and $\mu$
becomes $O(10)\%$ level.
We perform a Monte Carlo simulation for some minimal supergravity
benchmark points to demonstrate the detectability.
\end{abstract}


\preprint{}

\maketitle

\section{Introduction}

Supersymmetry (SUSY) is one of the promising candidates of the physics
beyond the standard model.
No signature of SUSY has been found yet.
However, the discovery of the SUSY particles is guaranteed for
$m_{\tilde{q}}, m_{\tilde{g}} \simle 2.5$ TeV at the CERN Large Hadron Collider
(LHC) for the minimal supergravity (mSUGRA) model.
Date taking is expected to start from 2007.
Masses of the sparticles will also be measured at the LHC \cite{LHC} with
reasonable accuracy, which is important to distinguish various SUSY
breaking models. 

The Lagrangian of supersymmetric theory is highly constrained. 
For example the sfermion-fermion-gaugino interaction is restricted to be of the
form $\tilde{f}_{L(R)}$-$f_{L(R)}$-$\tilde{G}$ because $\tilde{f}$ and
$f$ belong to the same chiral multiplet in supersymmetry.
On the other hand, $\tilde{f}_L$ and $\tilde{f}_R$ mixing terms are
controlled by the $F$ term of the superpotential of the form
$W = \mu H_1 H_2 + y_u Q U^c H_2 + y_d Q D^c H_1 + y_l L E^c H_1$.
Therefore the mixing term is proportional to the Yukawa coupling $y_f$
and $\mu$ parameter.
Studying such mixing terms determined by the supersymmetry would be an
interesting target for collider physics after the discovery of the
supersymmetry.

In this paper we study the effect of the neutralino polarization in the
decay chain
\begin{equation}
\tilde{q}\to q \tilde{\chi}^0_2 \to q l^\pm_1 \tilde{l}^\mp \to
q l^\pm_1 l^\mp_2 \tilde{\chi}^0_1,
\label{eq:decay}
\end{equation}
here $l_1(l_2)$ denotes the lepton from $\tilde{\chi}^0_2(\tilde{l})$ decay.
Due to the chirality structure of the squark-quark-neutralino coupling,
the $\tilde{\chi}^0_2$ is polarized.
The polarization of $\tilde{\chi}^0_2$ then affects the angular
distribution of the slepton in the
$\tilde{\chi}^0_2 \to l^\pm_1 \tilde{l}^\mp$ decay.
The polarization dependence of the angular distribution eventually shows
up in the charge asymmetry in the $m(q l^\pm_1)$ distribution, because
the polarization dependent part of the amplitude flips under the charge
conjugation transformation.
The effect can be seen at the $pp$ collider LHC because the number of
squarks $N(\tilde{q})$ produced at the LHC is larger than $N(\tilde{q}^*)$.
The polarization effect is considered in
Refs.~\cite{Richardson:2001df,Barr:2004ze},
where the slepton is assumed to be purely right-handed.
We study this process taking account of the left-right mixing of the
sleptons, which depends on the lepton flavor.
We point out that the left-right mixing significantly affects the charge
asymmetry as well as the decay width of
$\tilde{\chi}^0_2 \to l^\pm_1 \tilde{l}^\mp$ for $l=\mu$ and $\tau$.

This paper is organized as follows.
In the next section, we first summarize the relevant formulae and
discuss the effect of the left-right mixing qualitatively, then we
evaluate the lepton flavor dependence of the charge asymmetry and the
decay widths for explicit examples.
In Sec.~\ref{sec:simulation}, we carry out the Monte Carlo (MC) study and
discuss the feasibility to study the chiral effect at the LHC.
Sec.~\ref{sec:discussion} is devoted for discussions.

\section{Squark cascade decay and lepton charge asymmetry}

\subsection{Formalism}

We parameterize the sfermion-fermion-neutralino interaction Lagrangian as
\begin{eqnarray}
  {\cal L} &=& -\frac{ g_2 }{ \sqrt{2} }
  \sum_{i,\alpha}
  \bar{\tilde{\chi}}^0_i
  \left(
      L_{i\alpha}^{f}\frac{1-\gamma_5}{2}
    + R_{i\alpha}^{f}\frac{1+\gamma_5}{2}
  \right)
  f \tilde{f}^*_\alpha
  + {\rm H.c.},
\end{eqnarray}
where $i=1,2,3,4$ and $\alpha=1,2$ are the suffices for the mass
eigenstates of the neutralinos and the sfermions, respectively.
$g_2$ is the SU(2) gauge coupling constant.
For the charged lepton sector, the coupling constants $L_{i\alpha}^{l}$
and $R_{i\alpha}^{l}$ are given as
\begin{eqnarray}
  L_{i1}^{l} &=& 
  - \left[ (U_N^*)_{i2} + (U_N^*)_{i1}\tan\theta_W \right] \cos\theta_l
  + \frac{ m_l }{ m_W \cos\beta }(U_N^*)_{i3} \sin\theta_l,
\label{eq:L1}
\\
  L_{i2}^{l} &=& 
    \left[ (U_N^*)_{i2} + (U_N^*)_{i1}\tan\theta_W \right] \sin\theta_l
  + \frac{ m_l }{ m_W \cos\beta }(U_N^*)_{i3} \cos\theta_l,
\label{eq:L2}
\\
  R_{i1}^{l} &=&
    2(U_N)_{i1} \tan\theta_W \sin\theta_l,
  + \frac{ m_l }{ m_W \cos\beta }(U_N)_{i3} \cos\theta_l
\label{eq:R1}
\\
  R_{i2}^{l} &=&
    2(U_N)_{i1} \tan\theta_W \cos\theta_l,
  - \frac{ m_l }{ m_W \cos\beta }(U_N)_{i3} \sin\theta_l
\label{eq:R2}
\end{eqnarray}
where $U_N$ is the unitary matrix which diagonalize the neutralino mass
matrix.
The left-right mixing angle of the sleptons $\theta_l$ is defined as
follows.
The slepton mass term is written as
\begin{eqnarray}
  -{\cal L}_{\textrm{mass}}( \tilde{l} ) &=&
  \left(
    \begin{array}{cc}
      \tilde{l}_L^* & \tilde{l}_R^*
    \end{array}
  \right)
  {\cal M}_{\tilde{l}}^2
  \left(
    \begin{array}{c}
      \tilde{l}_L \\ \tilde{l}_R
    \end{array}
  \right),
\end{eqnarray}
where the mass squared matrix ${\cal M}_{\tilde{l}}^2$ is given as
\begin{eqnarray}
  {\cal M}_{\tilde{l}}^2 &=&
  \left(
    \begin{array}{cc}
        m_{\tilde{l}_L}^2 + m_l^2
        + m_Z^2\cos2\beta(\sin^2\theta_W-\frac{1}{2})
      & -m_l( A_{l}^* + \mu\tan\beta) \\
        -m_l( A_{l} + \mu^*\tan\beta)
      & m_{\tilde{l}_R}^2 + m_l^2
        - m_Z^2\cos2\beta\sin^2\theta_W
    \end{array}
  \right).
\label{eq:msl}
\end{eqnarray}
In this paper, we neglect lepton flavor mixing and assume that $A_l$ and
$\mu$ are real.
Then $\theta_l$ is obtained by diagonalizing the slepton mass matrix:
\begin{eqnarray}
  U {\cal M}_{\tilde{l}}^2 U^\dagger &=& 
  \left(
    \begin{array}{cc}
      m^2_{\tilde{l}_1} & 0 \\ 0 & m^2_{\tilde{l}_2}
    \end{array}
  \right),
~~~
U =
 \left(
   \begin{array}{cc}
     \cos\theta_{l} & \sin\theta_{l} \\
     -\sin\theta_{l} & \cos\theta_{l}
   \end{array}
 \right),
\label{eq:thetal}
\end{eqnarray}
where we assume $m^2_{\tilde{l}_1}<m^2_{\tilde{l}_2}$.
The mass eigenstates $\tilde{l}_{1,2}$ are defined as
\begin{eqnarray}
 \left(
   \begin{array}{c}
     \tilde{l}_1 \\ \tilde{l}_2
   \end{array}
 \right)
 &=&
 U
 \left(
   \begin{array}{c}
     \tilde{l}_L \\ \tilde{l}_R
   \end{array}
 \right).
\end{eqnarray}
The coupling constants for the squark sector are obtained in the same
way.

The decay widths of the subprocesses
$\tilde{f}_\alpha \to f \tilde{\chi}^0_i$ and
$\tilde{\chi}^0_i \to l^\pm \tilde{l}^\mp_\alpha$ are written as
\begin{eqnarray}
  \Gamma( \tilde{f}_\alpha \to f \tilde{\chi}^0_i )
  &=&
  \frac{ g_2^2 m_{\tilde{f}_\alpha} }{ 32\pi }
  \left(
    1 - \frac{ m_{\tilde{\chi}^0_i}^2 }{ m_{\tilde{f}_\alpha}^2 }
  \right)^2
  \left(
    |L_{i\alpha}^{f}|^2 + |R_{i\alpha}^{f}|^2
  \right),
\\
  \Gamma( \tilde{\chi}^0_i \to l^\pm \tilde{l}_\alpha^\mp )
  &=&
  \frac{ g_2^2 m_{\tilde{\chi}^0_i} }{ 64\pi }
  \left(
    1 - \frac{ m_{\tilde{l}_\alpha}^2 }{ m_{\tilde{\chi}^0_i}^2 }
  \right)^2
  \left(
    |L_{i\alpha}^{l}|^2 + |R_{i\alpha}^{l}|^2
  \right).
\end{eqnarray}
Here and hereafter, we neglect the quark and lepton masses for the
kinematics, while keeping them in the coupling constants
Eqs.~(\ref{eq:L1})--(\ref{eq:R2}).

The momentum configuration of the whole decay chain
$\tilde{q}_\beta\to q \tilde{\chi}^0_2 \to q l^\pm_1 \tilde{l}^\mp_\alpha
\to q l^\pm_1 l^\mp_2 \tilde{\chi}^0_1$ is described by three angular
variables.
The distribution of the decay chain is given as
\begin{eqnarray}
  \frac{ d^3\Gamma }{ d\cos\theta_{\tilde{l} }
    d\cos\theta_{\tilde{\chi}^0_1} d\phi_{\tilde{\chi}^0_1} }
  &=&
  \frac{1}{8\pi} \Gamma(\tilde{q}_\beta \to q \tilde{\chi}^0_2 )
  Br(\tilde{\chi}^0_2 \to l^\pm_1 \tilde{l}^\mp_\alpha )
  Br(\tilde{l}^\mp_\alpha \to l^\mp_2 \tilde{\chi}^0_1 )
  \nonumber\\&&\times
  \left[
    1 \mp A(l)\cos\theta_{\tilde{l}}
  \right],
\label{eq:dist}
\\
  A(l) &=&
  \frac{ |L_{2\beta}^{q}|^2 - |R_{2\beta}^{q}|^2 }
       { |L_{2\beta}^{q}|^2 + |R_{2\beta}^{q}|^2 }
  \cdot
  \frac{ |L_{2\alpha}^{l}|^2 - |R_{2\alpha}^{l}|^2 }
       { |L_{2\alpha}^{l}|^2 + |R_{2\alpha}^{l}|^2 },
\label{eq:asym}
\end{eqnarray}
where $Br$ denotes the branching ratio
$Br(\tilde{\chi}^0_2 \to l^\pm \tilde{l}^\mp_\alpha)
=\Gamma(\tilde{\chi}^0_2 \to l^\pm \tilde{l}^\mp_\alpha)
/\Gamma_{\textrm{total}}$, {\it etc}.
$\theta_{\tilde{l}}$ is the angle between the momenta of the quark and
the lepton $l_1$ in the $\tilde{\chi}^0_2$ rest frame,
$\theta_{\tilde{\chi}^0_1}$ is the angle between the two lepton momenta
in the slepton rest frame, and $\phi_{\tilde{\chi}^0_1}$ is the angle
between the decay planes of $\tilde{q} \to q l^\pm_1 \tilde{l}^\mp$ and
$\tilde{\chi}^0_2 \to l^\pm_1 l^\mp_2 \tilde{\chi}^0_1$.
Since squark and slepton decays are spherically symmetric in the rest
frames of the decaying particles, the angular distribution is flat over
$\cos\theta_{\tilde{\chi}^0_1}$ and $\phi_{\tilde{\chi}^0_1}$.
The $\theta_{\tilde{l}}$ dependence comes from the polarization of
$\tilde{\chi}^0_2$ and the chirality structure of the
sfermion-fermion-neutralino interaction shows up in this angular
distribution.
Qualitatively if $\tilde{q}_L$ decays into $\tilde{\chi}^0_2$ followed
by $\tilde{\chi}^0_2 \to l^+ \tilde{l}^-$, the lepton favors going in the same
(opposite) direction to $\tilde{\chi}^0_2$ for $\tilde{l}_{L(R)}$.

Instead of the above angular variables, we consider the invariant masses of
the combinations of  the quark and the leptons, which are directly
connected as the observables.
The relations are given as
\begin{eqnarray}
  m^2(l_1 l_2) &=&
  \frac{
    ( m_{\tilde{\chi}^0_2}^2 - m_{\tilde{l}}^2 )
    ( m_{\tilde{l}}^2 - m_{\tilde{\chi}^0_1}^2 )
  }{ m_{\tilde{l}}^2 }
  \frac{ 1 - \cos\theta_{\tilde{\chi}^0_1} }{ 2 },
\\
  m^2(q l_1) &=&
  \frac{
    ( m_{\tilde{q}}^2 - m_{\tilde{\chi}^0_2}^2 )
    ( m_{\tilde{\chi}^0_2}^2 - m_{\tilde{l}}^2 )
  }{ m_{\tilde{\chi}^0_2}^2 }
  \frac{ 1 - \cos\theta_{\tilde{l}} }{ 2 },
\\
  m^2(q l_2) &=&
  \frac{ 1 }{ 4 }
  \frac{
    ( m_{\tilde{q}}^2 - m_{\tilde{\chi}^0_2}^2 )
    ( m_{\tilde{l}}^2 - m_{\tilde{\chi}^0_1}^2 )
  }{ m_{\tilde{\chi}^0_2}^2 \, m_{\tilde{l}}^2 }
\nonumber\\&&\times  
  \left[
    m_{\tilde{\chi}^0_2}^2
    ( 1 + \cos\theta_{\tilde{l}} )
    ( 1 - \cos\theta_{\tilde{\chi}^0_1} )
    +
    m_{\tilde{l}}^2
    ( 1 - \cos\theta_{\tilde{l}} )
    ( 1 + \cos\theta_{\tilde{\chi}^0_1} )
  \right.
\nonumber\\&&\phantom{\times}
  \left.
    + 2 m_{\tilde{\chi}^0_2} m_{\tilde{l}}
    \sin\theta_{\tilde{l}} \sin\theta_{\tilde{\chi}^0_1}
    \cos\phi_{\tilde{\chi}^0_1}
  \right].
\end{eqnarray}
The polarization effect of $\tilde{\chi}^0_2$ can be seen clearly in the
charge asymmetry in the $m(q l_1^\pm)$ distribution, since an
$A(l)$ is the coefficient of $\theta_{\tilde{l}}$, and it  
affects  the decay distributions for 
$\tilde{q}\to q \tilde{\chi}^0_2 \to q l_1^+ \tilde{l}^-$
and
$\tilde{q}\to q \tilde{\chi}^0_2 \to q l_1^- \tilde{l}^+$
with opposite signs.
However, the leptons $l_1$ of the neutralino decay are indistinguishable
from $l_2$ of the slepton decay.
We therefore study the charge asymmetry between the $m(ql^\pm)$
distributions taking both $l_1$ and $l_2$ into account.
The $m(ql^+)$ distribution is the sum of the $m(q l_1^+)$ 
distribution from the decay chain
$\tilde{\chi}^0_2 \to l_1^+\tilde{l}^- \to l_1^+l_2^-\tilde{\chi}^0_1$
and the $m(ql_2^+)$ distribution from
$\tilde{\chi}^0_2 \to l_1^-\tilde{l}^+ \to l_1^-l_2^+\tilde{\chi}^0_1$.

The charge asymmetry in $m(ql^\pm)$ distribution was studied in
Ref.~\cite{Barr:2004ze} in a case where the slepton is purely
right-handed: $L_{2\alpha}^{l}=0$.
In this paper we take account of the left-right mixing of the sleptons,
whose effect depends on the lepton flavor.

\subsection{Effect of the slepton left-right mixing in mSUGRA model}

We consider a ``typical'' case of the minimal supergravity model.
The SU(2) and U(1) gaugino masses $M_2$ and $M_1$, respectively, are
related to each other as $M_1 \approx 0.5 M_2$ and the magnitude of the
Higgsino mass $\mu$ is assumed to be larger than $M_2$, so that the wino
component dominates $\tilde{\chi}^0_2$ and the bino component dominates
$\tilde{\chi}^0_1$.
As for the slepton mass matrix, the right-handed slepton mass parameter
becomes smaller than the left-handed one due to the running effect.
Therefore the lighter slepton, $\tilde{l}_1$, is $\tilde{l}_R$-like and
the heavier one, $\tilde{l}_2$, is $\tilde{l}_L$-like in the most of the
parameter space.

We can safely neglect the left-right mixing of the squarks because the
process we consider is the decay of the first generation squark.
The first decay process $\tilde{q} \to q \tilde{\chi}^0_2$ in
Eq.~(\ref{eq:decay}) occurs predominantly through the
$\tilde{q}_L$-$q$-$\tilde{W}$ coupling if
$\tilde{\chi}^0_2 \sim \tilde{W}$.
Consequently the first factor of the right-hand side of
Eq.~(\ref{eq:asym}) is very close to unity.

Let us now consider the decay process
$\tilde{\chi}^0_2 \to l^\pm \tilde{l}^\mp_1$ with $l=e$, $\mu$, or
$\tau$.
We first note that, in the minimal supergravity model, the slepton
left-right mixing angle $\theta_l$ is approximately written as
\begin{eqnarray}
  \tan2\theta_{l} &\approx&
  -\frac{ m_l( A_l + \mu \tan\beta ) }{ 0.37 M_{1/2}^2 }
  \left(
    1 + O\left(\frac{m_0^2 m_l^2\tan^2\beta}{M_{1/2}^2 m_W^2}\right)
  \right),
\end{eqnarray}
where $m_0$ and $M_{1/2}$ are the universal scalar mass and the gaugino
mass at the GUT scale, respectively, and $A_l$ is the trilinear scalar
coupling constant at the low energy scale.
The numerical factor 0.37 in the denominator is determined by the gauge
coupling constants.
In the (s)electron case, the left-right mixing of the selectrons is
negligible, {\it i.~e.}, $\cos\theta_l\approx 0$ for $l=e$ in
Eqs.~(\ref{eq:L1})--(\ref{eq:R2}), and the charge asymmetry is maximal
($A(e)\approx -1$).

The effect of the left-right mixing and the Yukawa coupling is quite
significant for the (s)tau mode.
For $l=\tau$, the typical magnitude of $\cos\theta_{\tau}$ is $O(1)$ for
large $\tan\beta$ in mSUGRA.
Magnitudes of the neutralino mixing matrix elements are typically
$|(U_N)_{22}|\approx 1$ and $|(U_N)_{21,23}|=O(10^{-1})$.
Substituting these numbers in Eqs.~(\ref{eq:L1}) and (\ref{eq:R1}), we
see that $L_{21}^{\tau}$ can dominate over $R_{21}^{\tau}$ and the
behavior of the charge asymmetry is opposite to the electron case.

The left-right mixing effect may be observed even in the (s)muon case.
Since both terms of the right-hand side of Eq.~(\ref{eq:L1}) are
approximately proportional to the lepton mass, $L_{21}^{\mu}$ is
enhanced by $O(m_\mu/m_e)$ compared to $L_{21}^{e}$.
For relatively large $\tan\beta$, it is possible that
$L_{21}^{\mu}$ and $R_{21}^{\mu}$ are of the same order of magnitude, as
we will see later.

Notice that the branching ratio of $\tilde{\chi}^0_2 \to l \tilde{l}_1$ 
also depends on the lepton flavor.
The coupling $R_{21}^{l}$ is suppressed due to the small $(U_N)_{21}$
component, therefore even tiny mixing of the left-hand component will
have a significant effect.
Especially, the decay width for $\tau \tilde{\tau}_1$ mode is much
larger than those for $e \tilde{e}_1$ and $\mu \tilde{\mu}_1$, and
$\Gamma( \tilde{\chi}^0_2 \to \mu \tilde{\mu}_1) >
\Gamma( \tilde{\chi}^0_2 \to e \tilde{e}_1)$.

We also consider the case where $m_{\tilde{\chi}^0_2}>m_{\tilde{l}_2}$.
In this case, both $\tilde{l}_1$ and $\tilde{l}_2$ contribute to the
decay chain so that the decay distribution over $m(ql^\pm)$ is more
complicated.
Since $\tilde{l}_2$ is  $\tilde{l}_L$-like, the coupling $L_{22}^{l}$ in
Eq.~(\ref{eq:L2}) dominates the
$\tilde{\chi}^0_2 \to l^\pm \tilde{l}_2^\mp$ decay and the branching
ratio of $\tilde{\chi}^0_2 \to l^\pm \tilde{l}_2^\mp$ is much larger
than that of $\tilde{\chi}^0_2 \to l^\pm \tilde{l}_1^\mp$.
The main contribution of the charge asymmetry therefore comes from
$\tilde{l}_2$.

Let us study the lepton flavor dependence more quantitatively.
In order to compare with the Monte Carlo simulations in the next
section, we take benchmark points SPS1a and SPS3 given in
Ref.~\cite{Allanach:2002nj}, where the mass spectrum is calculated by
ISAJET \cite{Baer:1999sp}.
The model points we study are as follows;
\begin{itemize}
\item SPS1 \cite{Allanach:2002nj}:\\
This is the mSUGRA point where $m_0=100$~GeV, $M_{1/2}=250$~GeV, the
trilinear coupling at the GUT scale $A_0=-100$~GeV, $\tan\beta=10$ and
$\mu>0$.
The decay $\tilde{\chi}^0_2\rightarrow \tilde{l}_Rl$ is open.
We also show some results for the point where $m_0$ and $M_{1/2}$ are
the same as those of SPS1a but $\tan\beta=15$ and 20.
For these points, the effect of $\tilde{\mu}$ left-right mixing in
$\tilde{\chi}^0_2$ decay becomes significant.
\item  SPS3 (or point C of \cite{Battaglia:2001zp}):\\
$m_0=90$~GeV, $M_{1/2}=400$~GeV, $A_0=0$, $\tan\beta=10$ and $\mu>0$.
The decay $\tilde{\chi}^0_2\rightarrow \tilde{l}_L l$ is open for this
point.
The decay must show opposite charge asymmetry to that of $\tilde{l}_{R}$.
\end{itemize}
The kinematics of the decay distribution is important for 
our study. We therefore list the calculated end points 
for the decay (\ref{eq:decay}) in Table \ref{list}.
\begin{table}
\begin{tabular}{|c||c|c|c|c|c|c|c|c|}
\hline
model & $m_{\tilde{u}_L}$ &$m_{\tilde{\chi}^0_2}$
 & $m_{\tilde{\chi}^0_1}$ & $m_{\tilde{l}}$ & $m_{ql_1}$ & $m_{ql_2}$
 & $m_{qll}$ & $m_{ll}$
\\
\hline
SPS1a & 537.3 & 176.8 & 96.1 & 143.0 & 298.4 & 375.7 & 425.9 & 77.0
\\
SPS3($\tilde{l}_R$)
 & 832.4 & 303.7 & 164.0 & 182.3 & 619.9 & 310.2 & 652.3 & 106.1
\\
SPS3($\tilde{l}_L$)
 & & & & 292.8 & 208.8 & 642.0 & 652.3 & 66.8
\\
\hline
\end{tabular}
\caption{
The sparticle masses  and end points of $qll$ distributions arising from
the cascade decay
$\tilde{q} \to q\tilde{\chi}^0_2 \to ql_1\tilde{l}
\to ql_1l_2\tilde{\chi}^0_1$ at SPS1a and SPS3.
The masses and end points for SPS1a($\tan\beta=15$, 20) are  similar to
that for $\tan\beta=10$ for the first and second generation sfermions.
All masses are given in GeV.
}
\label{list}
\end{table}

Since the masses and the Yukawa couplings of the first and second
generations are neglected in ISAJET, its output cannot be used directly
for the discussion of $e$--$\mu$ difference.
We take account of the left-right mixings of selectrons and smuons in
the following way.
The output from ISAJET contains a set of the low energy MSSM parameters
$m_{\tilde{q}_L}$, $m_{\tilde{d}_R}$, $m_{\tilde{u}_R}$, 
$m_{\tilde{l}_L}$ and $m_{\tilde{e}_R}$
for the first generation sfermions, and
$m_{\tilde{q}_{3L}}$, $m_{\tilde{b}_R}$, $m_{\tilde{t}_R}$, 
$m_{\tilde{l}_{3L}}$, $m_{\tilde{\tau}_R}$,
$A_{t}$, $A_{b}$ and $A_{\tau}$
for the third generation sfermions,
as well as the parameters in the gaugino and Higgs(ino) sectors.
The soft SUSY breaking parameters for the second generation sfermions
are assumed to be the same as the first generation ones.
Since the trilinear couplings $A_{e,\mu}$ are not given in the ISAJET
output, we take $A_{e}=A_{\mu}=A_{\tau}$, but the left-right mixing
mainly comes from the $\mu\tan\beta$ term anyway.
Then we diagonalize the mass matrix (\ref{eq:msl}) to obtain the mixing
angle $\theta_{l}$ in (\ref{eq:thetal}) and evaluate the coupling
constants $L_{i\alpha}^{l}$ and $R_{i\alpha}^{l}$.
The slepton mass eigenvalues obtained with this procedure are close to
the original ISAJET output.
The numerical values of the coupling constants relevant to the
$\tilde{\chi}^0_2 \to l \tilde{l}_{\alpha}$ are given in
Table~\ref{tab:LR}.
>From these values, we obtain
\begin{eqnarray}
  \Gamma(e) : \Gamma(\mu) : \Gamma(\tau)
  &\approx&
  1 : 1.04 : 13.7,
\\
  A(e) : A(\mu) : A(\tau)
  &\approx&
  1 : 0.93 : -0.84,
\end{eqnarray}
for the SPS1a.
Here $\Gamma(l)$ denotes the decay width
$\Gamma(\tilde{\chi}^0_2 \to l^\pm \tilde{l}_1^\mp)$.
\begin{table}[htbp]
\centering
\begin{tabular}{cccccc}
\hline
Model & $l$ & $|L_{21}^{l}|$ & $|R_{21}^{l}|$ & $|L_{22}^{l}|$ & $|R_{22}^{l}|$
\\
\hline
SPS1a & $e$ & $1.06\times10^{-4}$ & 0.116 & 0.999 & $2.81\times10^{-5}$
\\
 & $\mu$ & 0.0219 & 0.116 & 0.998 & $5.81\times10^{-3}$
\\
 & $\tau$ & 0.327 & 0.0950 & 0.946 & 0.0909
\\
\hline
SPS3 & $e$ & $6.47\times10^{-5}$ & 0.0598 & 0.990 & $1.77\times10^{-5}$
\\
 & $\mu$ & 0.0134 & 0.0598 & 0.990 & $3.66\times10^{-3}$
\\
 & $\tau$ & 0.215 & 0.0505 & 0.968 & 0.0603
\\
\hline
\end{tabular}
\caption{
Magnitude of slepton-lepton-neutralino coupling constants for the
benchmark points SPS1a and SPS3.
}
\label{tab:LR}
\end{table}

We see that the effect of the smuon left-right mixing is rather small in
the above model points, where $\tan\beta=10$.
However, $L_{21}^{l}$ is enhanced for larger $\tan\beta$ since the
source of the slepton left-right mixing is the Yukawa coupling, while
$R_{21}^{l}$, whose main component is the U(1) gauge coupling, is less
sensitive to $\tan\beta$.
Therefore, we expect a larger $e$--$\mu$ difference in $\Gamma(l)$ and
$A(l)$ for larger $\tan\beta$.
In order to see the $\tan\beta$ dependence, we consider variants of
SPS1a, where the mSUGRA parameters have the same values as those in
SPS1a, but $\tan\beta$ is taken to be different.
The relative magnitudes of the decay widths and $A(l)$, as well as the
magnitudes of relevant coupling constants for $\tan\beta=10$, 15 and 20
are listed in Table~\ref{tab:tanbeta}.
We see $|L_{21}^{\mu}|$ is proportional to $\tan\beta$ in this region
and the $e$--$\mu$ differences in $\Gamma(l)$ and $A(l)$ are quite
significant for $\tan\beta=20$.
\begin{table}
\begin{tabular}{cccccc}
\hline
$\tan\beta$ & $l$ & $|L_{21}^{l}|$ & $|R_{21}^{l}|$ &
$\Gamma(l)/\Gamma(e)$ & $A(l)/A(e)$
\\
\hline
10 & $e$ & $1.06\times10^{-4}$ & 0.116 & 1 & 1
\\
 & $\mu$ & 0.0219 & 0.116 & 1.04 & 0.93
\\
 & $\tau$ & 0.327 & 0.0950 & 13.7 & $-0.84$
\\
\hline
15 & $e$ & $1.58\times10^{-4}$ & 0.107 & 1 & 1
\\
 & $\mu$ & 0.0326 & 0.107 & 1.09 & 0.83
\\
 & $\tau$ & 0.438 & 0.0671 & 37.6 & $-0.95$
\\
\hline
20 & $e$ & $2.09\times10^{-4}$ & 0.102 & 1 & 1
\\
 & $\mu$ & 0.0432 & 0.102 & 1.17 & 0.70
\\
 & $\tau$ & 0.522 & 0.0419 & 80.2 & $-0.99$
\\
\hline
\end{tabular}
\caption{
Absolute values of the coupling constants and relative magnitudes of the
decay widths and $A(l)$ for different choices of $\tan\beta$.
The other mSUGRA parameters are same as those in SPS1a.
}
\label{tab:tanbeta}
\end{table}

Besides the slepton-lepton-neutralino coupling constants, the mass
spectrum is also affected by changing the value of $\tan\beta$.
However, since the shifts are tiny for $m_{\tilde{e}_1}$,
$m_{\tilde{\mu}_1}$ and $m_{\tilde{\chi}^0_2}$, the $\tan\beta$
dependence of the $e$--$\mu$ difference in the decay width and the
charge asymmetry comes dominantly from the $\tan\beta$ dependence of the
$\tilde{\mu}$ left-right mixing angle.
On the other hand, the mass eigenvalue of $\tilde{\tau}_1$ changes
substantially: 133 GeV for $\tan\beta=10$ to 108 GeV for $\tan\beta=20$,
namely.
Consequently, the $e$--$\tau$ differences and the $\tan\beta$
dependence in Table~\ref{tab:tanbeta} are the cumulative effect of both the
mixing angle and the mass eigenvalue.

\section{Simulations}
\label{sec:simulation}
\subsection{Charge asymmetry}

In this section, we show some simulation results at SPS1a and SPS3 and
discuss the LHC potential to study the chiral nature of sfermions at the
LHC in the decay (\ref{eq:decay}).

We generated $3\times 10^6$ events for SPS1a($\tan\beta=10$) and SPS3.
This corresponds to $\int{\cal L}dt=58~{\rm fb}^{-1}$ for SPS1a and
$600~{\rm fb}^{-1}$ for SPS3 respectively.
The mass spectrum and branching ratios are generated by ISAJET
\cite{Baer:1999sp} and interfaced to HERWIG \cite{Corcella:2000bw},
which means the effect of the non-universal branching ratios of
$\tilde{\chi}^0_2\to \tilde{e}e$ and $\tilde{\mu}\mu$ discussed in the
previous section is not taken into account in our event generations.
The events are studied using the fast detector simulator ATLFAST
\cite{atlfast}.
Cuts for the SUSY event selection are listed in Table \ref{cutlist}.
%
\begin{table}
\begin{tabular}{l}
$n_j(P_T>100$~GeV$)\geq 1$\\
$n_j(P_T>50$~GeV$)\geq 4$\\
$n_{\rm b \ jet}(P_T>50)=0$\\
$m_{\rm eff}>600$~GeV, where $m_{\rm eff}=p^T_{\rm miss} +p^T_{j1}
+p^T_{j2}+p^T_{j3}+p^T_{j4}$
\\
$E_T({\rm miss})>0.2 m_{\rm eff}$\\
Two and only two opposite sign isolated leptons \\
$P_{Tl_1}>20$~GeV,  $P_{Tl_2}>10 $~GeV\\
\end{tabular}
\caption{
Selection cuts used for the simulations in this paper.
}
\label{cutlist}
\end{table}

The dominant production processes are $pp\to \tilde{g}\tilde{g}$ and
$\tilde{g}\tilde{q}$ at the LHC.
The production cross section for the first generation squarks are larger
than that of anti-squarks, because the LHC is a $pp$ collider.
On the other hand, a gluino decay is charge symmetric, therefore the
squarks and anti-squarks from gluino decays dilute the total charge
asymmetry of SUSY signal.
For this simulation, we especially veto the events containing $b$ jets
with $P_T>50$~GeV
as they dominantly come from the decays $\tilde{g}\to \tilde{b}b$ and
$\tilde{t}{t}$.
For the points studied in this paper, more than $40\%$ of the gluino
decay involves $b$ jets.
After applying the cuts in Table \ref{cutlist}, the number of events
which involves $\tilde{u}_L$ or $\tilde{d}_L$ direct production are
50(58)\% of the total events for SPS1a(SPS3) respectively.

We have not performed a detailed simulation of the Standard Model
background for this study. The applied cuts are however very similar
to the cuts applied in \cite{giacomo}, where the SM background is shown 
to be negligible with respect to the SUSY background.
For the SPS3 Point, for which the signal statistics is lower and
we perform the study for an integrated luminosity of 600 pb$^{-1}$,
the SM background could be approximately at the level of the 
SUSY background. The subtraction procedure used for the SUSY 
background does work for the SM background which is 
dominated by $\bar tt$ production. Therefore 
we expect no significant modification of the result for SPS1a,
and only a moderate reduction of the significance for SPS3, 
due to increased statistical fluctuation.

\begin{figure}
\begin{center}
\includegraphics[height=5cm,clip]{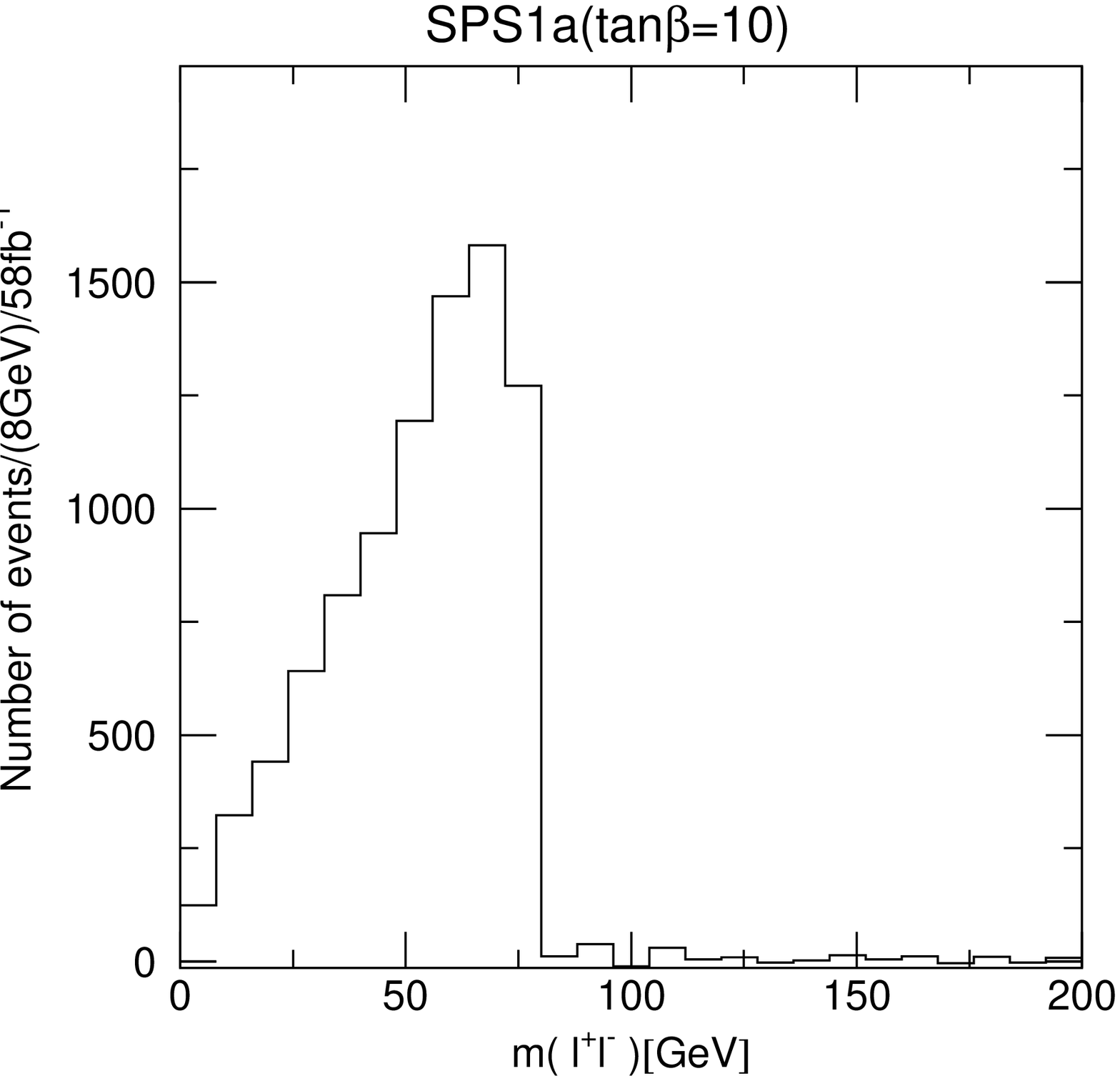}
\includegraphics[height=5cm,clip]{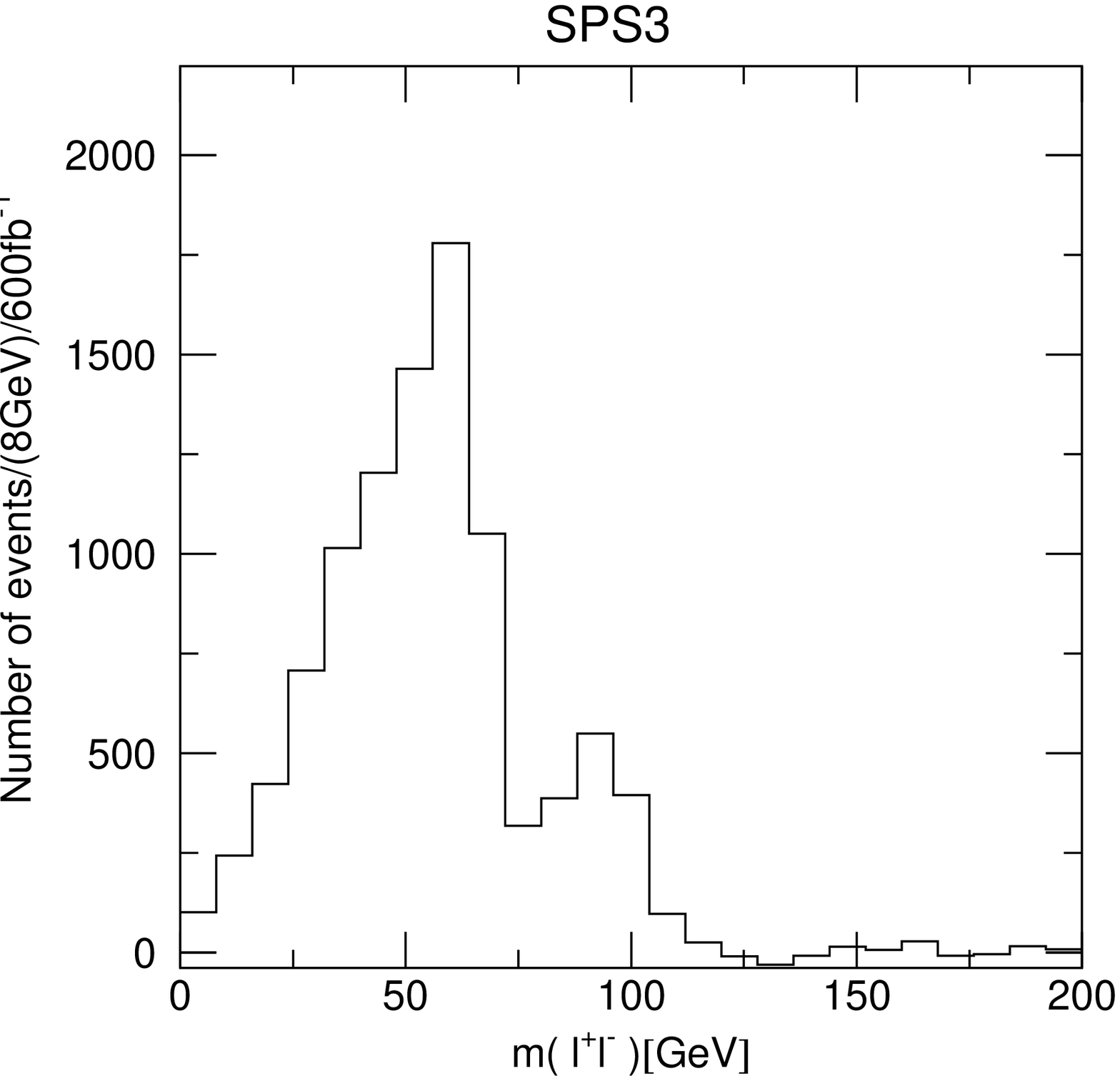}
\end{center}
\begin{center}
a)\hskip 5cm b)
\end{center}
\caption{
$m(ll)$ distributions for a) SPS1a and b) SPS3.
}
\label{mll}
\end{figure}

\begin{figure}
\includegraphics[height=5cm,clip]{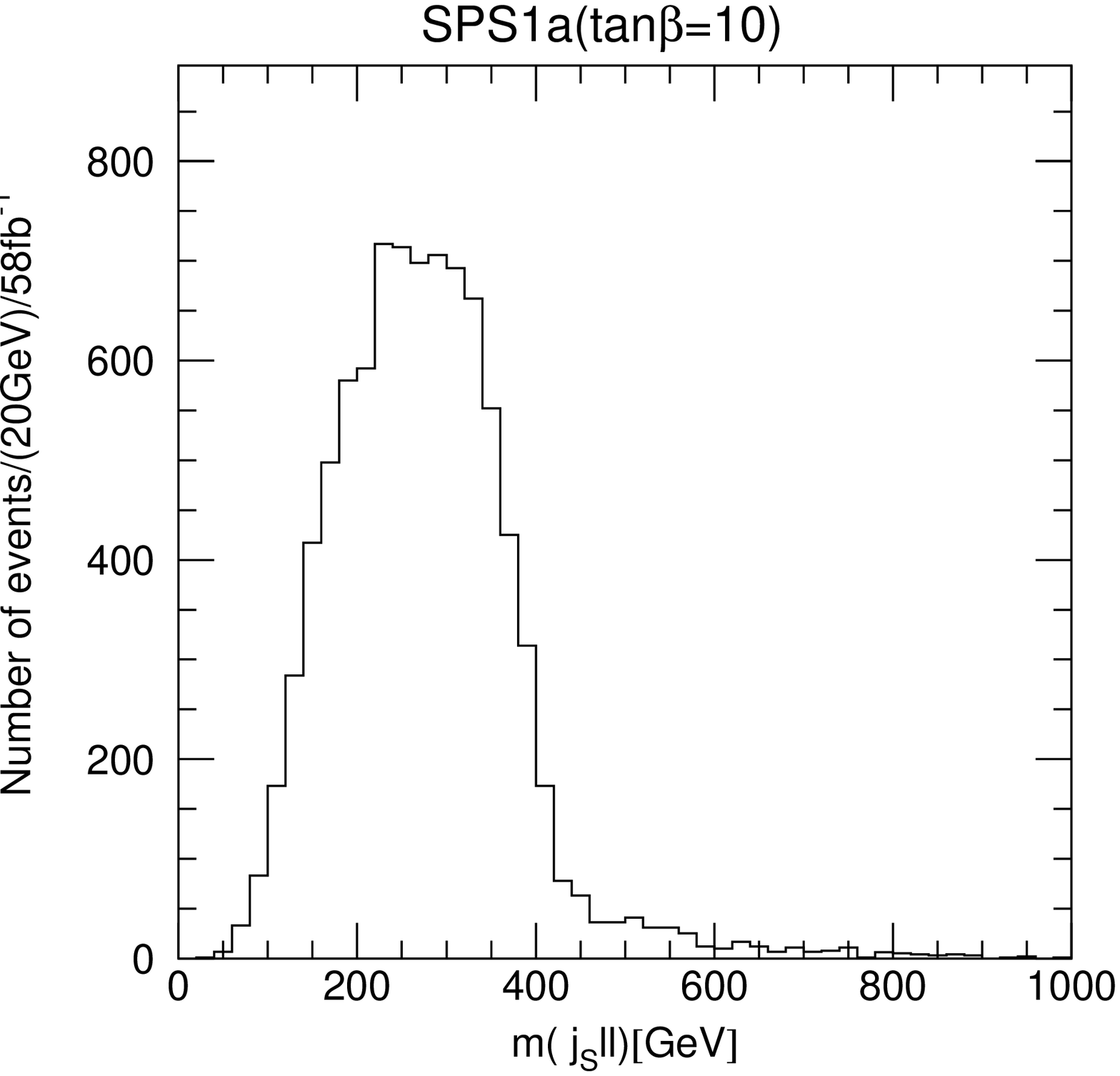}
\includegraphics[height=5cm,clip]{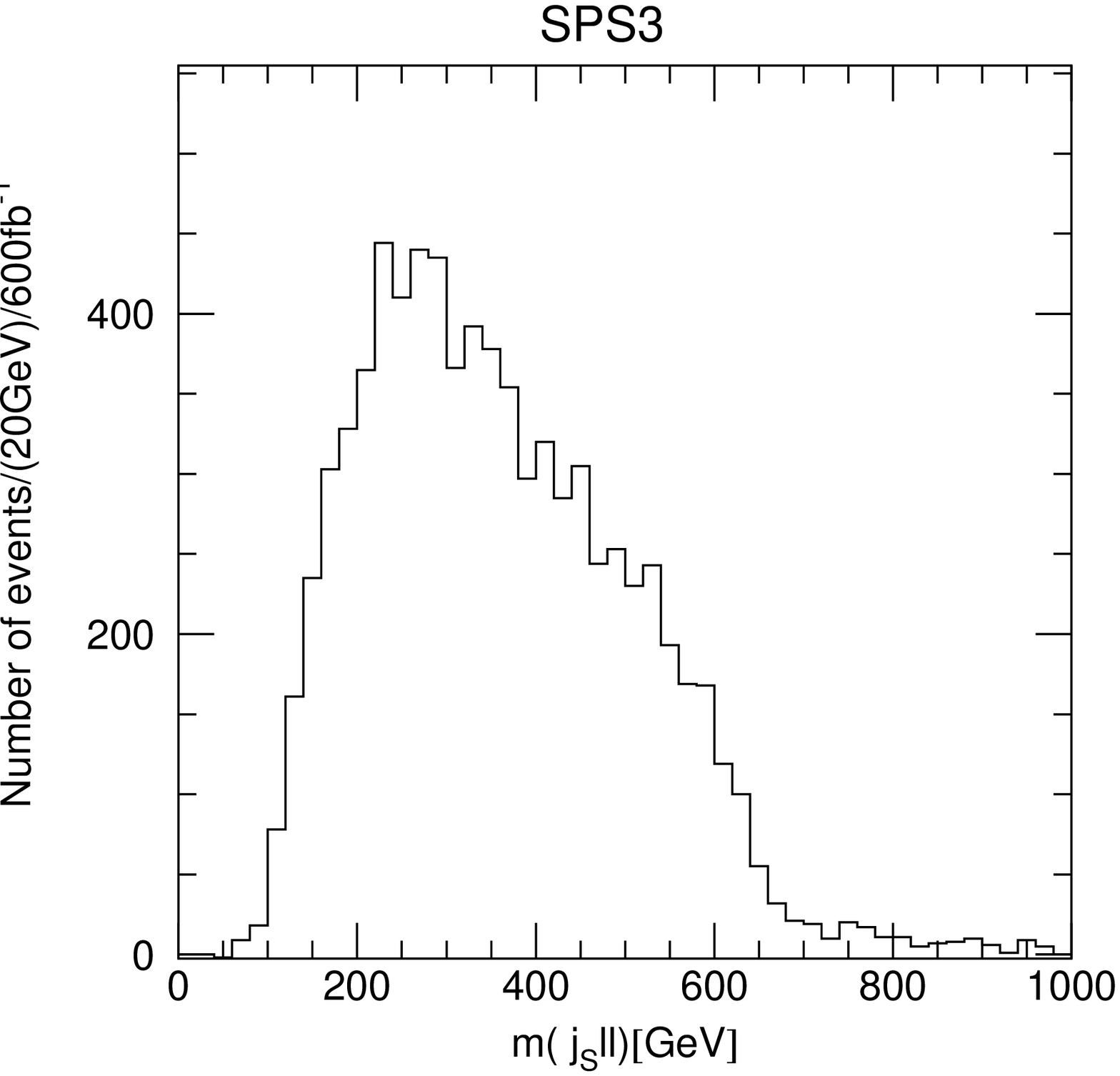}
\begin{center}
a)\hskip 5cm b) 
\end{center}
\caption{The $m(j_Sll)$ distributions for a) SPS1a and b) SPS3.}
\label{mjll}
\end{figure}

In Fig.~\ref{mll} and \ref{mjll}, we show some of the $j l^+l^-$
distribution at SPS 1a($\tan\beta=10$) and SPS3.
In Fig.~\ref{mll} we show $m(ll)$ distributions, where we subtract the
distributions of the events with odd sign odd flavor leptons (OSOF,
$e^+\mu^-$ and $\mu^+e^-$) from the events with odd sign same flavor
leptons (OSSF, $e^+e^-$ and $\mu^+\mu^-$).
The subtraction is known as a way to remove the backgrounds where two
leptons are originating from $W$ or $\tau$ decays.
Different detection efficiencies of electrons and muons may affect 
the subtraction, generating a systematic uncertainty in the 
result. The evaluation of these uncertainties is a complex
experimental task, outside the scope of this study. We will therefore
in the following only quote statistical errors.
We can however note that the experiments 
will be able to exploit a very large sample of 
$W \to l\nu$ and $Z\to l^+l^-$ decays to understand in detail the
acceptance and efficiency for  electrons and muons.

The $m(ll)$ distribution for the decay Eq.~(\ref{eq:decay}) has an edge
structure at $m(ll,{\rm max})$, which  are reconstructed for both of
the model points.
Especially two edges are seen for SPS3, and the lighter one is
consistent with the lepton pairs from the decay
$\tilde{\chi}^0_2\to\tilde{l}_L$, $m(ll,{\rm max}) =66.8$~GeV.

We study the distribution of the two leptons and one of the 
two highest $p_T$ jets, as one of the two jets likely comes
from $\tilde{q}$ decays if the mass difference between 
$\tilde{q}$ and $\tilde{\chi}^0_2$ is large enough. 
We denote the two jets as $j_S$ and 
$j_L$ where $m(j_S ll)<m(j_L ll)$.
The clear endpoints are seen 
by taking  $m(j_Sll)$ distributions as can be seen in Fig~\ref{mjll}.

\begin{figure}
\begin{center}
\includegraphics[height=5cm,clip]{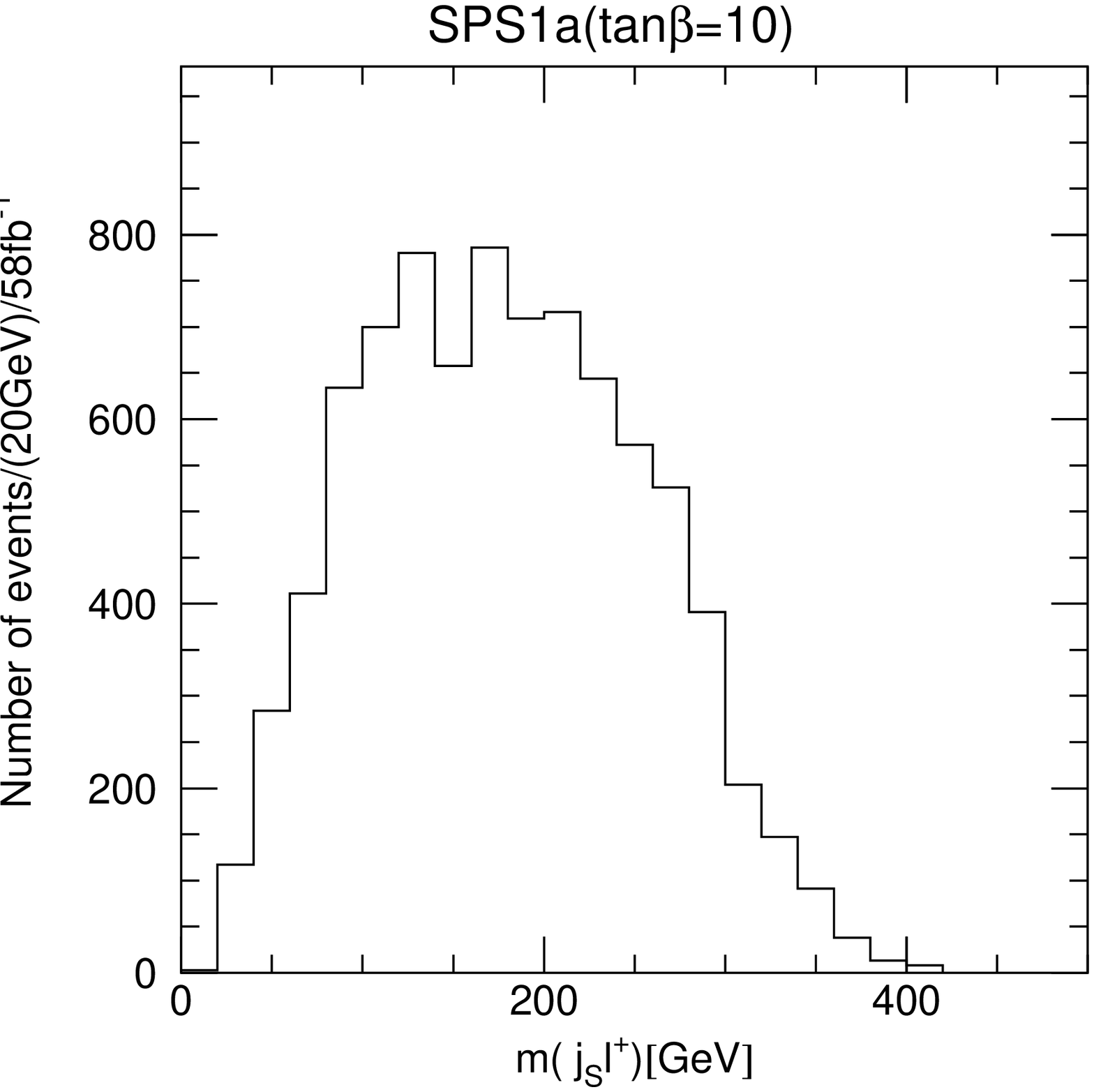}
\includegraphics[height=5cm,clip]{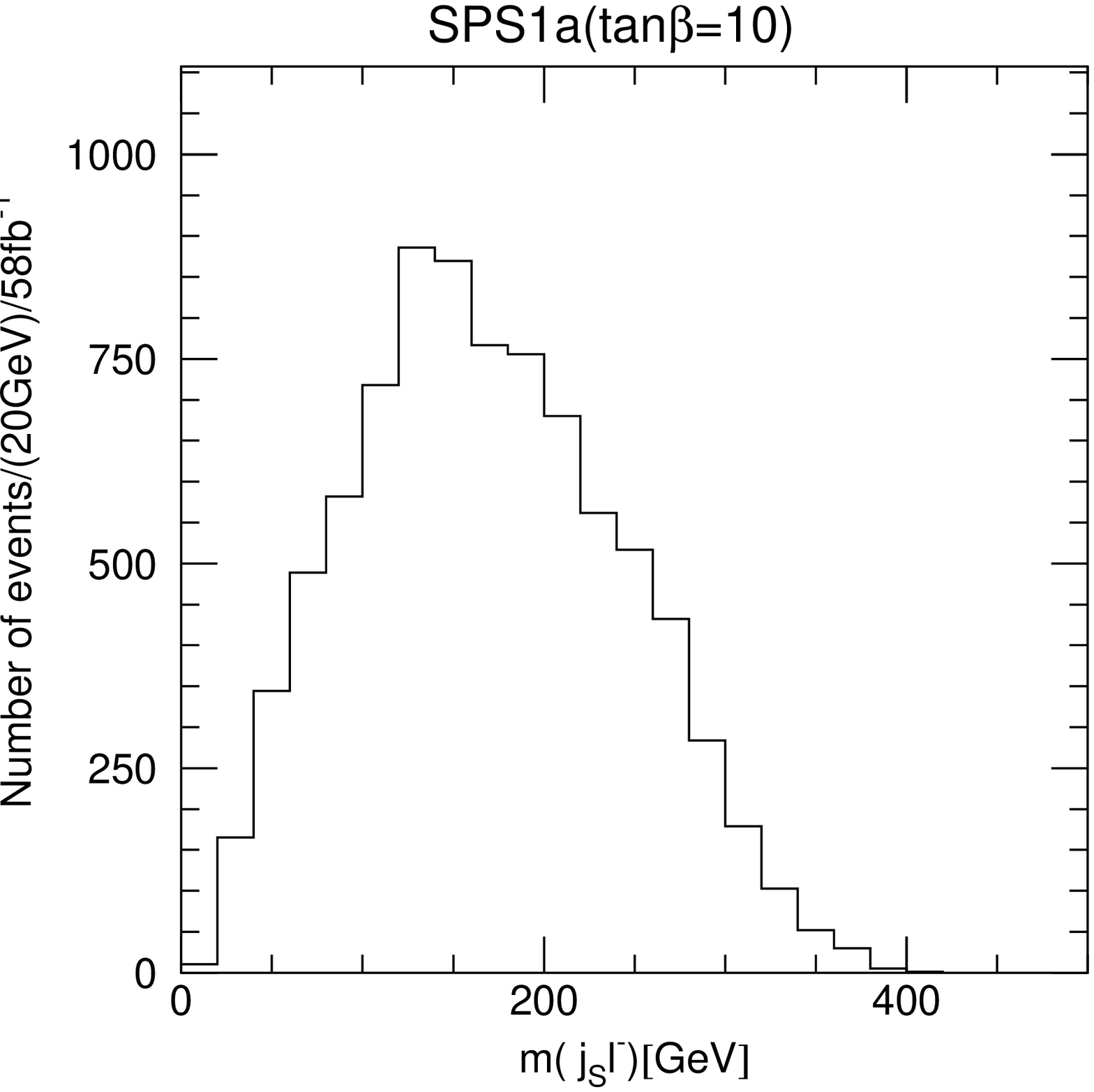}
\\
\includegraphics[height=5cm,clip]{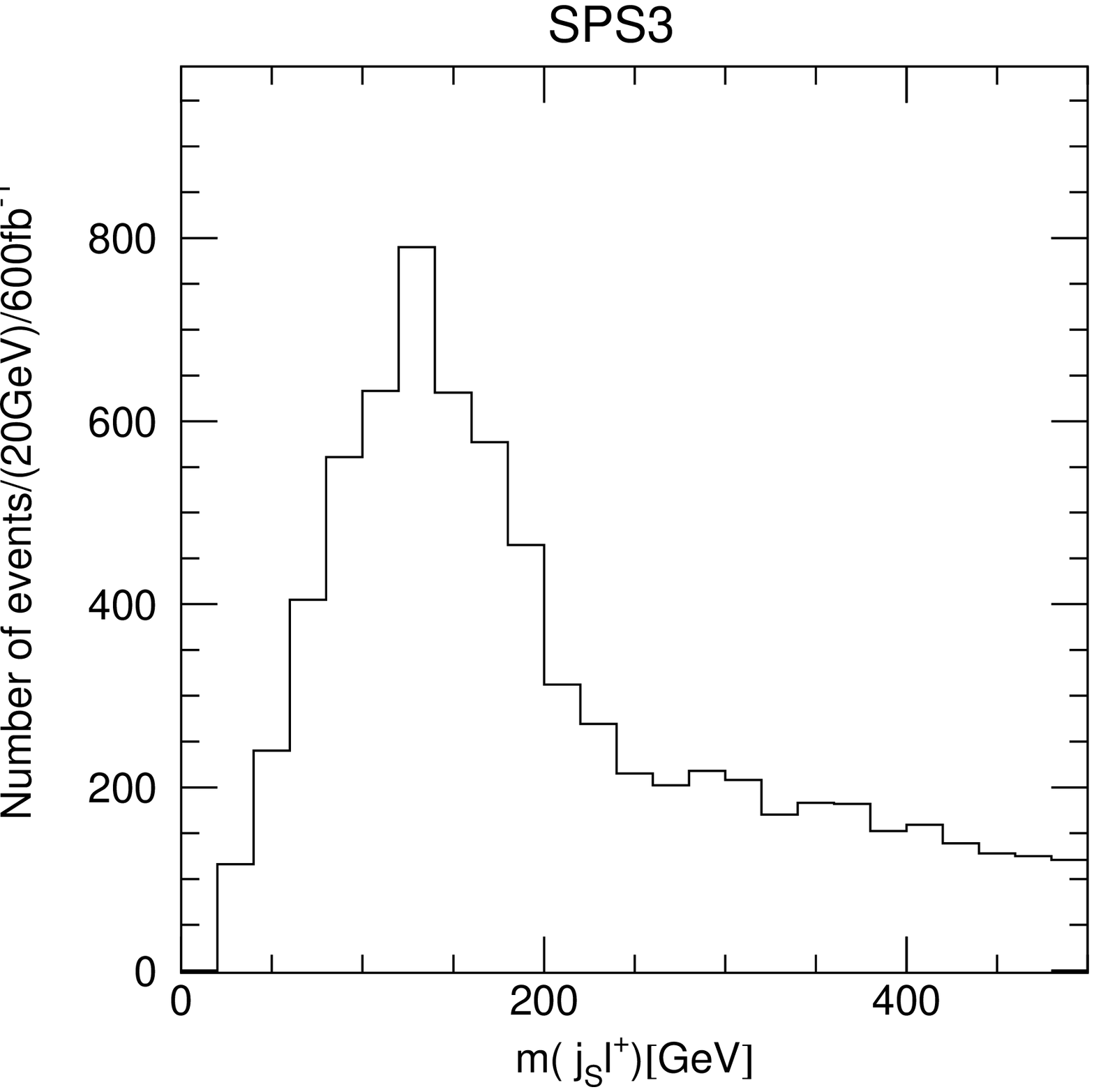}
\includegraphics[height=5cm,clip]{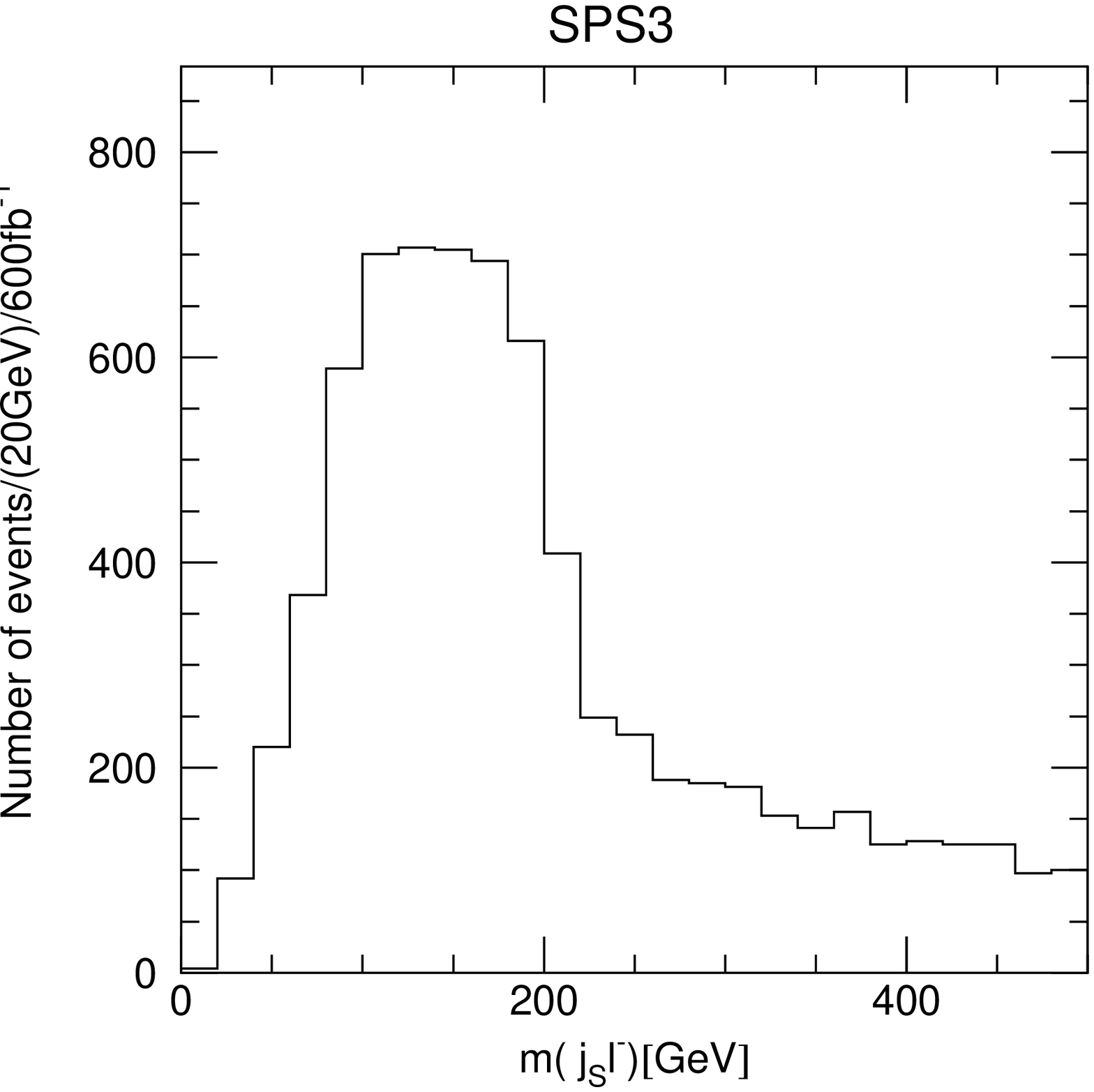}
\\
a)\hskip 5cm  b) 
\end{center}
\caption{
a) $m(j_Sl^+)$ and b) $m(j_Sl^-)$ distributions for SPS1a (the upper
plots) and SPS3 (the lower plots).
}
\label{mjsl}
\end{figure}

The $m(j_Sl)$ distributions are shown in Fig.~\ref{mjsl} a) and b).  
When the decay Eq.~(\ref{eq:decay}) is open, the $m(ql_1)$ distribution
has an edge.
The edge is very close to $m(jl,{\rm max})$ for SPS1a, while it is 
much lower than $m(jl,{\rm max})$ for SPS3, see Table \ref{list}. 
This is because 
$(m_{\tilde{\chi}^0_2}-m_{\tilde{l}})/m_{\tilde{\chi}^0_2}\ll 1$ 
for SPS3, therefore the lepton from $\tilde{\chi}^0_2$ decay is less
energetic.
For SPS1a (SPS3) the edge structure is clearly seen in 
$j_S l^+$ ($j_Sl^-$) distribution.
This is because the dominant mode is $l_R$($l_L$) for SPS1a (SPS3),
therefore $l^+$($l^-$) tend to  go to the opposite direction to $q$ in
the rest frame of $\tilde{q}$, respectively.

\begin{figure}
\begin{center}
\includegraphics[height=5cm,clip]{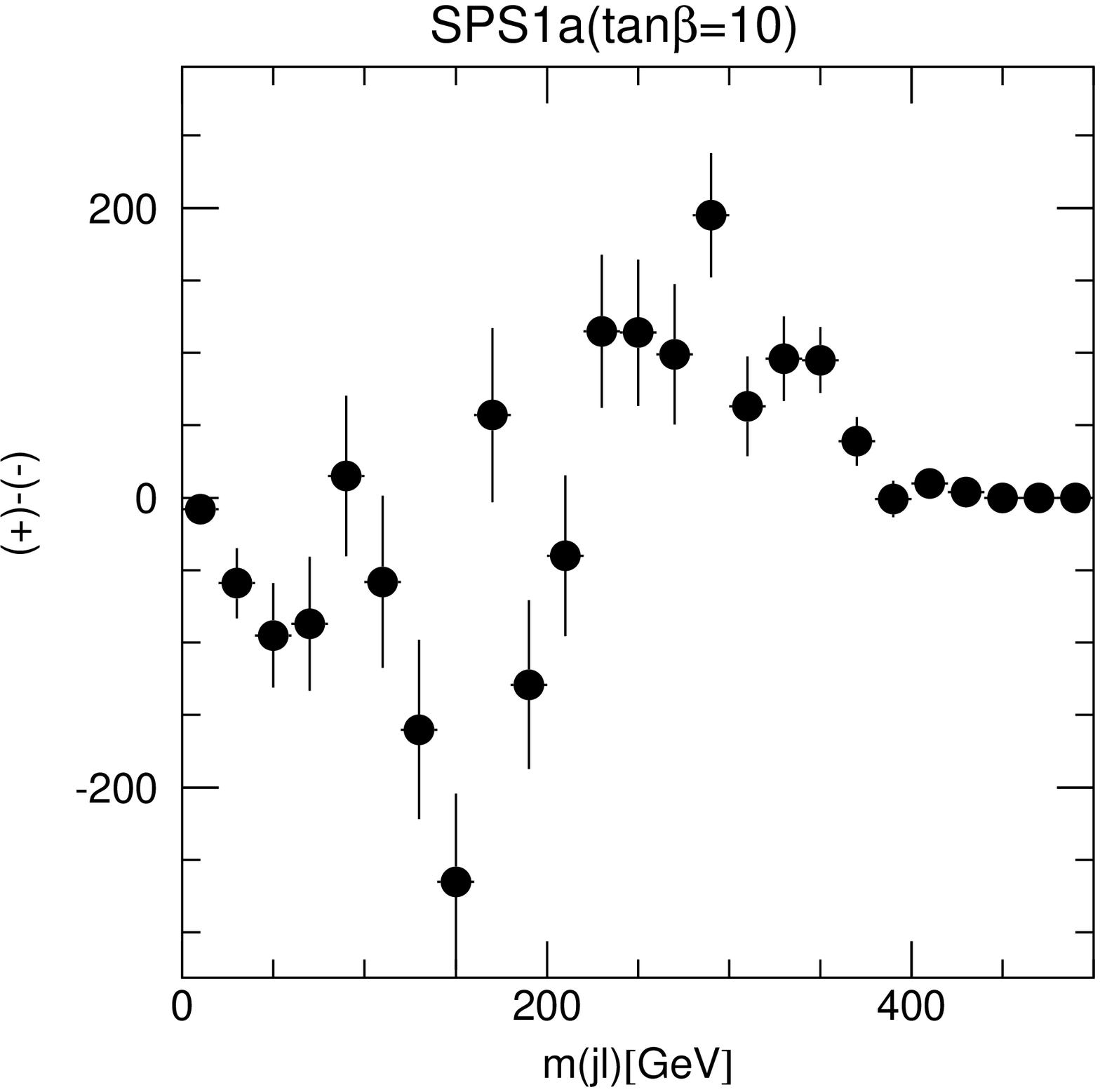}
\includegraphics[height=5cm,clip]{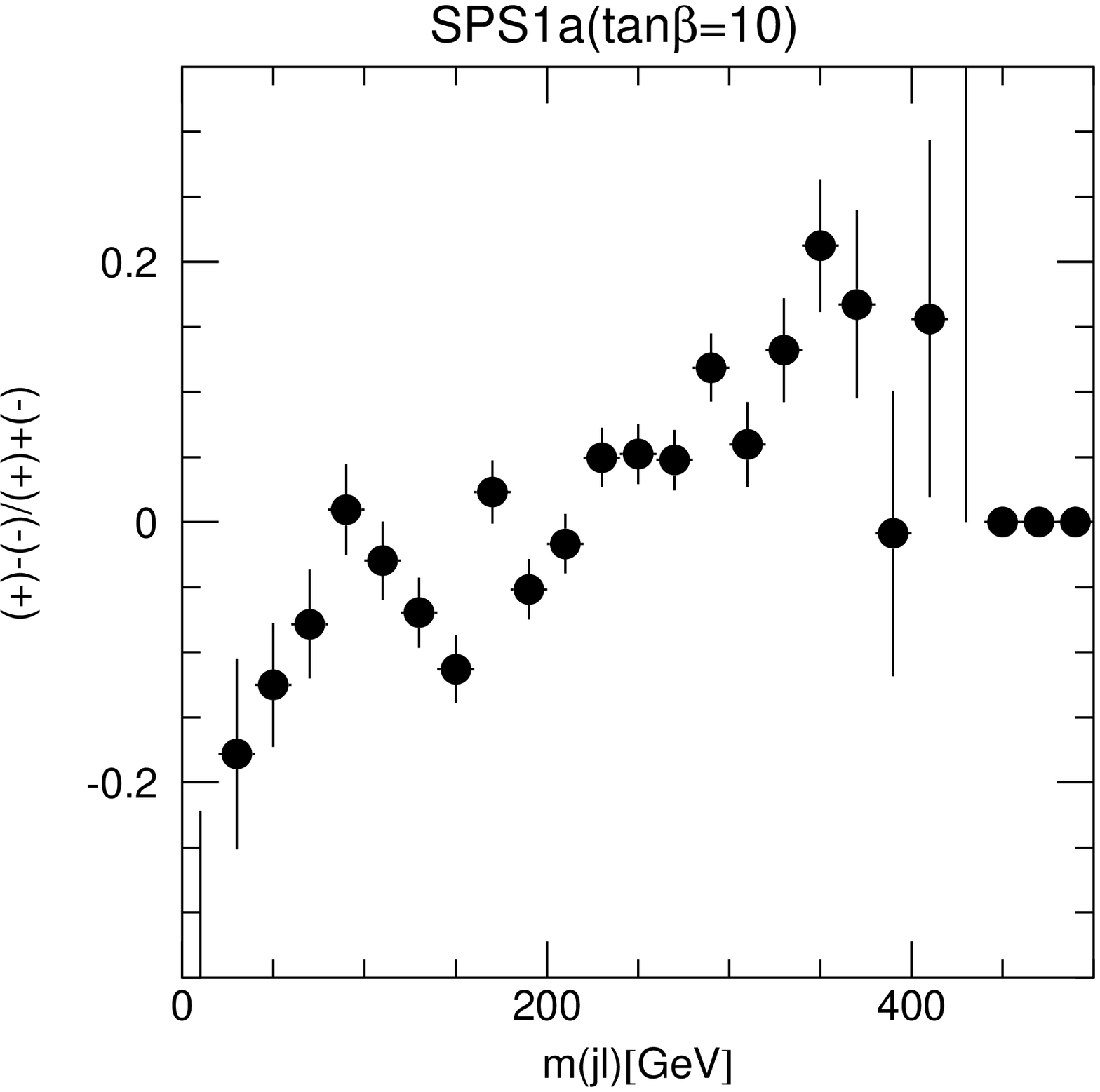}
\includegraphics[height=4.5cm,clip]{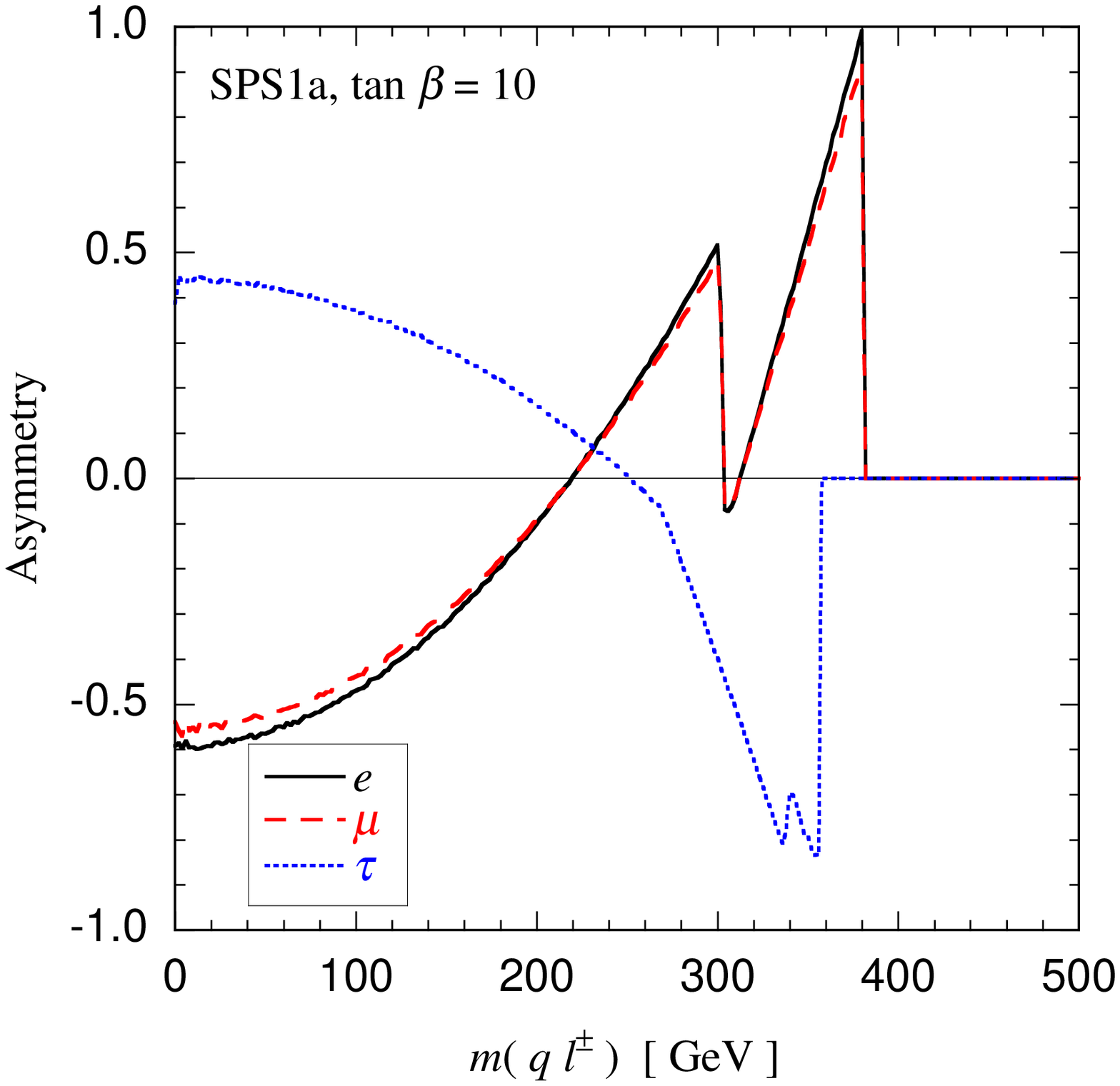}
\\
\includegraphics[height=5cm,clip]{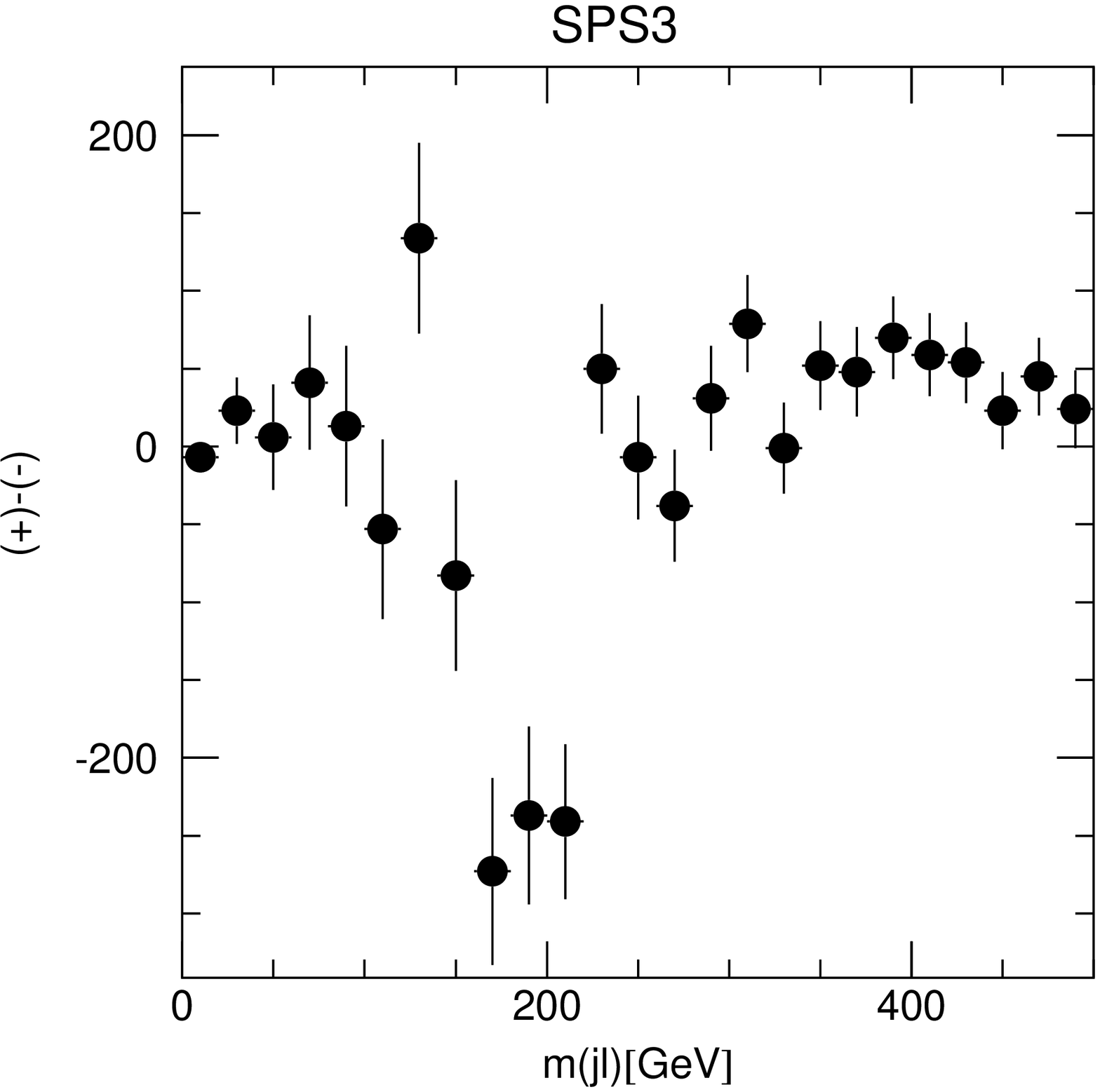}
\includegraphics[height=5cm,clip]{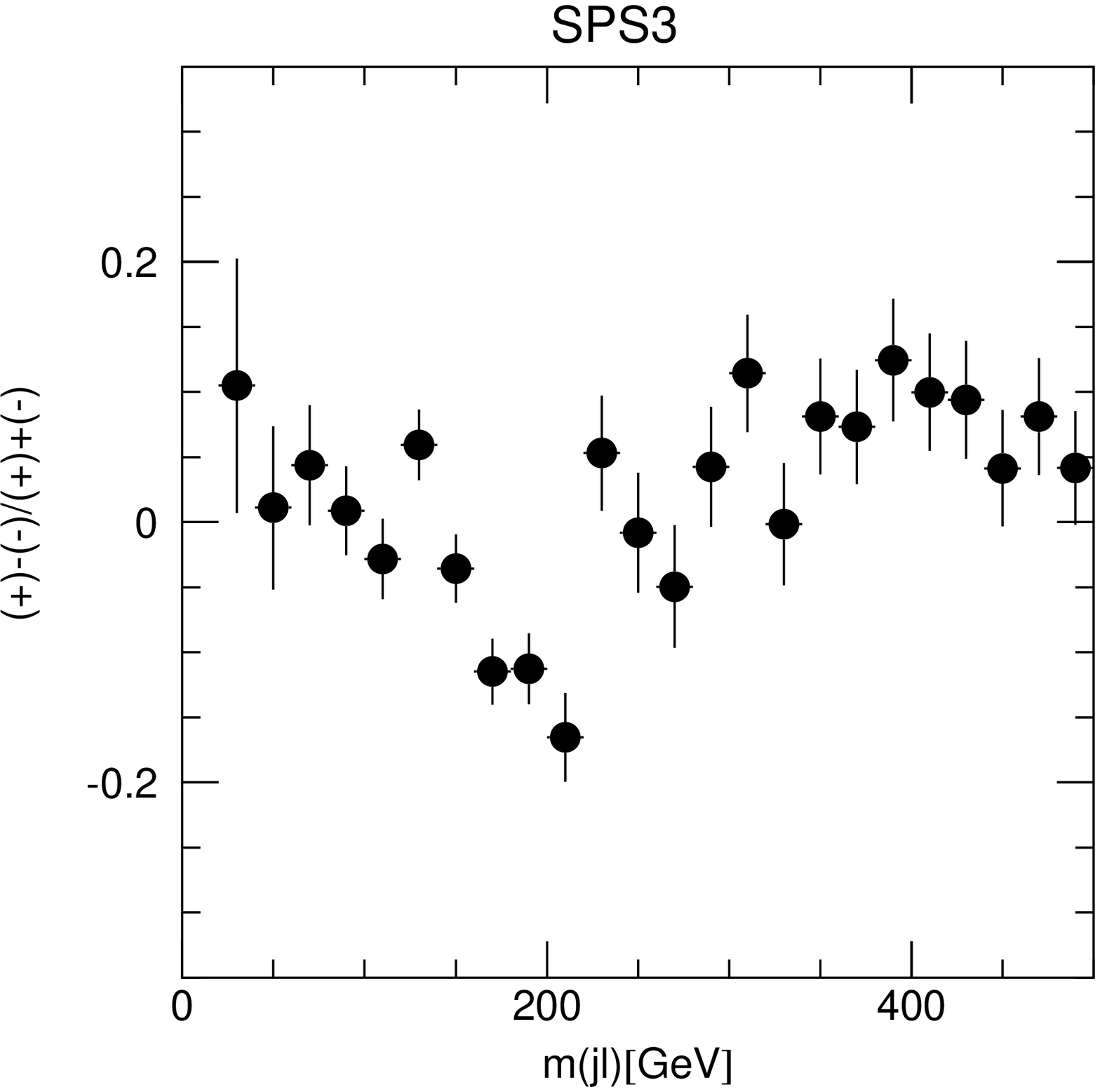}
\includegraphics[height=4.5cm,clip]{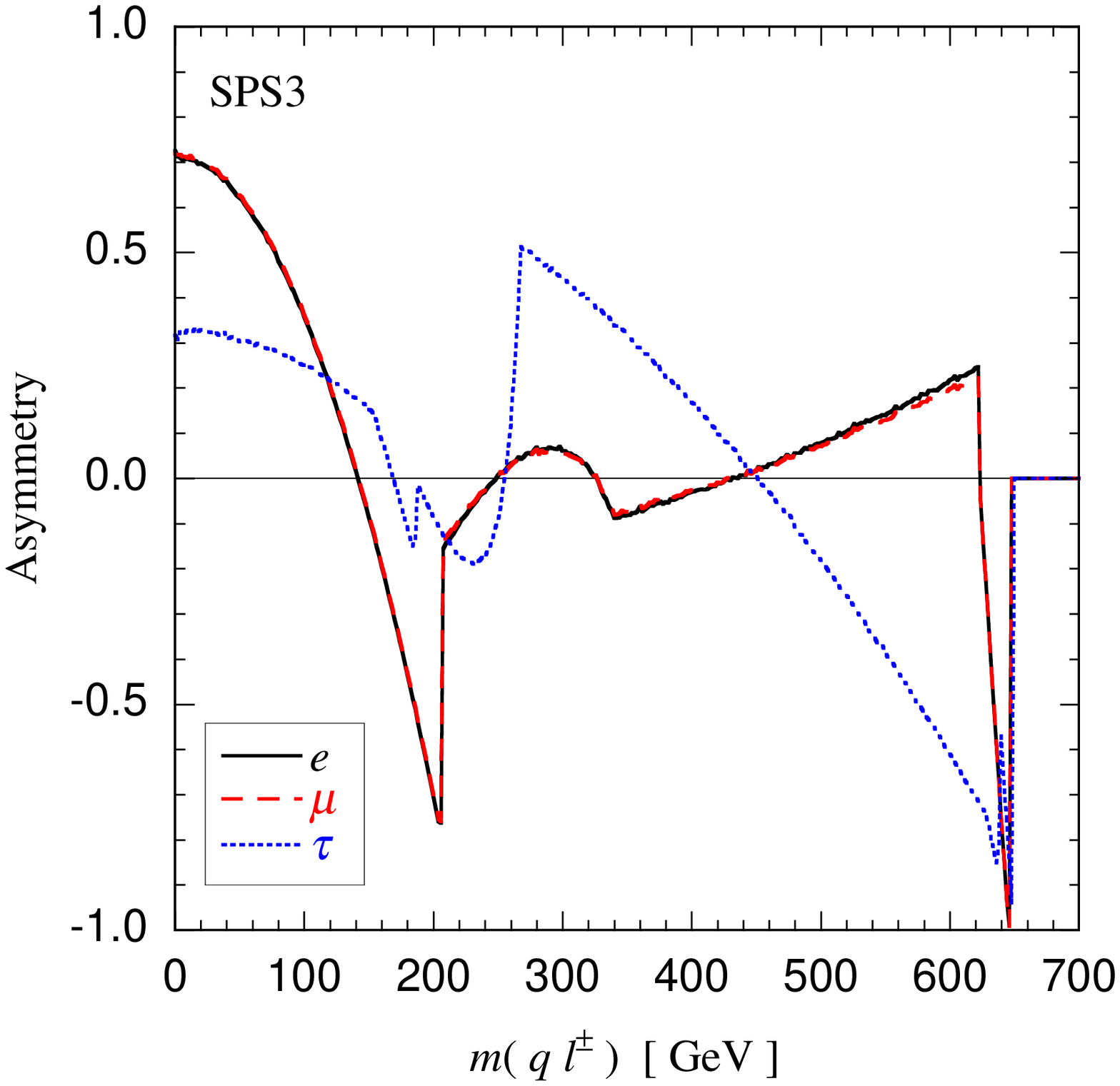}
\\
a)\hskip 5cm b) \hskip 5cm c)
\end{center}
\caption{
a) The difference between $m(jl^+)$ and  $m(jl^-)$ distributions. 
b) The reconstructed asymmetry 
$[N_{\rm sig}(jl^+)-N_{\rm sig}(jl^-)]
/[N_{\rm sig}(jl^+)+N_{\rm sig}(jl^-)]$.
c) The calculated $[N(ql^+)-N(ql^-)]/[N(ql^+)+N(ql^-)]$ for
$\tilde{q} \to q\tilde{\chi}^0_2 \to ql^{\mp}\tilde{l}^{\pm}
\to ql^{\mp}l^{\pm}\tilde{\chi}^0_1$ for $e^+e^-$, $\mu^+\mu^-$ and 
$\tau^+\tau^-$.
The differences for $j=j_S$ and $j=j_L$ are added.
For $j=j_L$ we require $m(j_Lll)<m(qll, {\rm max})$.
}
\label{asymmetry}
\end{figure}

Fig.~\ref{asymmetry} a) shows the difference 
of the number of events after the background subtractions 
$N(jl^{\pm})$ as the function of $m(jl^{\pm})$. 
The deviation from the zero are clearly seen.
We also show b) the reconstructed charge asymmetry
$[N_{\rm sig}(jl^+)-N_{\rm sig}(jl^-)]
/[N_{\rm sig}(jl^+)+N_{\rm sig}(jl^-)]$ as a function of 
$m(jl)$ and c) the calculated charge asymmetry for $m(ql)$. 
The distribution of the measured asymmetry is diluted due to the
$\tilde{q}^*$ productions and the decay and there are also the 
contamination of the events where a wrong jet is selected as $j_S$.
However b) and c) are qualitatively similar. 
The distribution shows positive asymmetry near $m(jl)$ endpoint 
for SPS1a, while it shows negative asymmetry for the smaller 
endpoint of $m(jl)$ distributions for SPS3. 

\begin{figure}
\begin{center}
\includegraphics[height=5cm,clip]{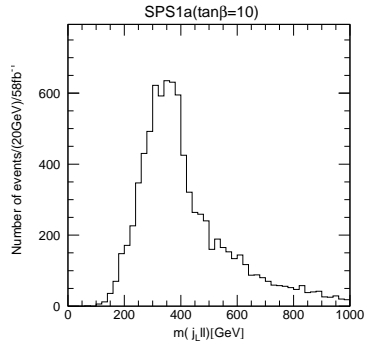}
\end{center}
\caption{The $m(j_Lll)$ distribution at SPS1a.}
\label{mjlll}
\end{figure}

For these charge asymmetry plots, we add the difference for $m(j_Sl)$
and for $m(j_Ll)$ distributions, therefore each event is used twice for
our analysis.
This is not a problem because events which contribute to the asymmetry in
$m(j_Sl)$ distribution and in $m(j_Ll)$ distribution must be
statistically independent.
Note a $j_S$ is more likely to be a wrong jet when $m(j_Sll)$ is much
smaller than the end point.
A significant fraction of $j_L$'s also originates from $\tilde{q}$ decay.
This can be seen in the $m(j_Lll)$ distribution at SPS1a, which shows
a bump structure $\sim m(qll, {\rm max})=425$~GeV as in
Fig.~\ref{mjlll}. In this paper, 
we require $m(jll)<450$~(700) and 
and require $m(ll)<80$~(70)~GeV for SPS1a(SPS3) except 
Fig.~\ref{mll}, Fig.~\ref{mjll}, and Fig.~\ref{mjlll}.

The statistical error of the asymmetry must be estimated based on the
number of the odd sign leptons $N_{\rm OS}$ rather than the
number of events after $e \mu$ event subtraction $N_{\rm sig}$.
We can check the statistical significance of 
the charge asymmetry by calculating
$\Delta \chi^2/{\rm n.d.f.}$ from a constant distribution as
\begin{equation}
\Delta \chi^2 =\min_{{\rm for}\ c} \sum_{m(jl)\ {\rm bins}}
\frac{\left[N_{\rm sig}(j l^+)-N_{\rm sig}(jl^-)-c\right]^2}
{N_{\rm OS}(jl^+) + N_{\rm OS}(jl^-)},
\end{equation}
which is listed in Table \ref{number}. 

\begin{table}
\begin{tabular}{|c||c|c|c|c|c|c|c|}
\hline
model points & $m_{\rm cut}^{\rm (min)}$ & $m_{\rm cut}^{\rm (max)}$  &
 \multicolumn{2}{c|}{$N_{\rm OS}(jl^+)+N_{\rm OS}(jl^-)$}  &
 \multicolumn{2}{c|}{$N_{\rm sig}(jl^+)-N_{\rm sig}(jl^-)$}  &
$S$
\\
&&& $j_S$ & $j_L$ & $j_S$ & $j_L$ &($\Delta\chi^2/{\rm n.d.f.}$)
\\
\hline
SPS1a($\tan\beta=10$) & 220&380 &
\ \  5426 \ \ & 7988 &\ \ 454\ \ &362 & 7.37\\ 
($e^+e^-+\mu^+\mu^-$)&&&&&&&(84.9/15 for 60--380~GeV)
\\
\hline
SPS3 & 140&220 &
 7691 & 8139 &\  $-439$\ \ & $-394$ & 6.43\\
($e^+e^-+\mu^+\mu^-$)&&&&&&&(51.9/7 for 60--220~GeV)
\\
\hline
\hline
SPS1a($\tan\beta=20$) & 220&380 &
 3389 & 4574 & 174 & 99  & 3.3\\
($e^+e^-$)&&&&&&&(24.6/15 for 60--380~GeV)
\\
\hline
\end{tabular}
\caption{
The statistical significance $S$ and $\Delta \chi^2/{\rm n.d.f.}$ are 
estimated using the MC sample for SPS1a ($\tan\beta=10,20$) and SPS3.  For
SPS1a($\tan\beta=10$) and SPS3, $3\times 10^6$ events are used, while
$15\times 10^6$ events are used for SPS1a($\tan\beta=20$).
$N_{\rm OS }(jl^{\pm})$ is the number of events with  odd sign  leptons 
with $ m^{\rm
(min)}_{\rm cut}<m(jl^{\pm})<m^{\rm (max)}_{\rm cut}$, while $N_{\rm
sig}(jl^{\pm})$ are the number of events after $e\mu$ subtractions.
}
\label{number}
\end{table}

We define the statistical significance of the the charge asymmetry $S$
as
\begin{eqnarray}
S^2 &=&
\left.\frac{\left[N_{\rm sig}(j_Sl^+)-N_{\rm sig}(j_Sl^-)\right]^2}
{N_{\rm OS}(j_Sl^+)+N_{\rm OS}(j_Sl^-)}\right\vert
_{m^{\rm min}_{\rm cut}<m(j_Sl)<m^{\rm max}_{\rm cut}}
\nonumber\\
&&+\left.\frac{\left[N_{\rm sig}(j_Ll^+)-N_{\rm sig}(j_Ll^-)\right]^2}
{N_{\rm OS}(j_Ll^+)+N_{\rm OS}(j_Ll^-)}\right\vert
_{m^{\rm min}_{\rm cut}<m(j_Ll)<m^{\rm max}_{\rm cut}}.
\end{eqnarray} 
where one of $m_{\rm cut}$ is taken to the value of $m(jl)$ where the
charge asymmetry changes its sign.
We do not take into account the event outside the region
$m^{\rm min}_{\rm cut}<m(jl)<m^{\rm max}_{\rm cut}$, where the asymmetry
is opposite.
If we add the contribution from both positive and negative
asymmetry bins, we might overestimate the statistical significance
because the $jl^+$ and $jl^-$ distributions are of the same $jll$
sample.

We list the statistical significance 
$S$ for SPS1a($\tan\beta=10$) and SPS3 for the
sum of $ee$ and $\mu\mu$ events in Table \ref{number}. 
The significance is based on the number of events for
$\int{\cal L}dt \sim 60~{\rm fb}^{-1}$ for SPS1a($\tan\beta=10$) while it
corresponds to $600~{\rm fb}^{-1}$ for SPS3.
For $300~{\rm fb}^{-1}$ the left hand nature of slepton at SPS3 may still
be seen with $S>3$.

\begin{figure}
\begin{center}
\includegraphics[height=5cm,clip]{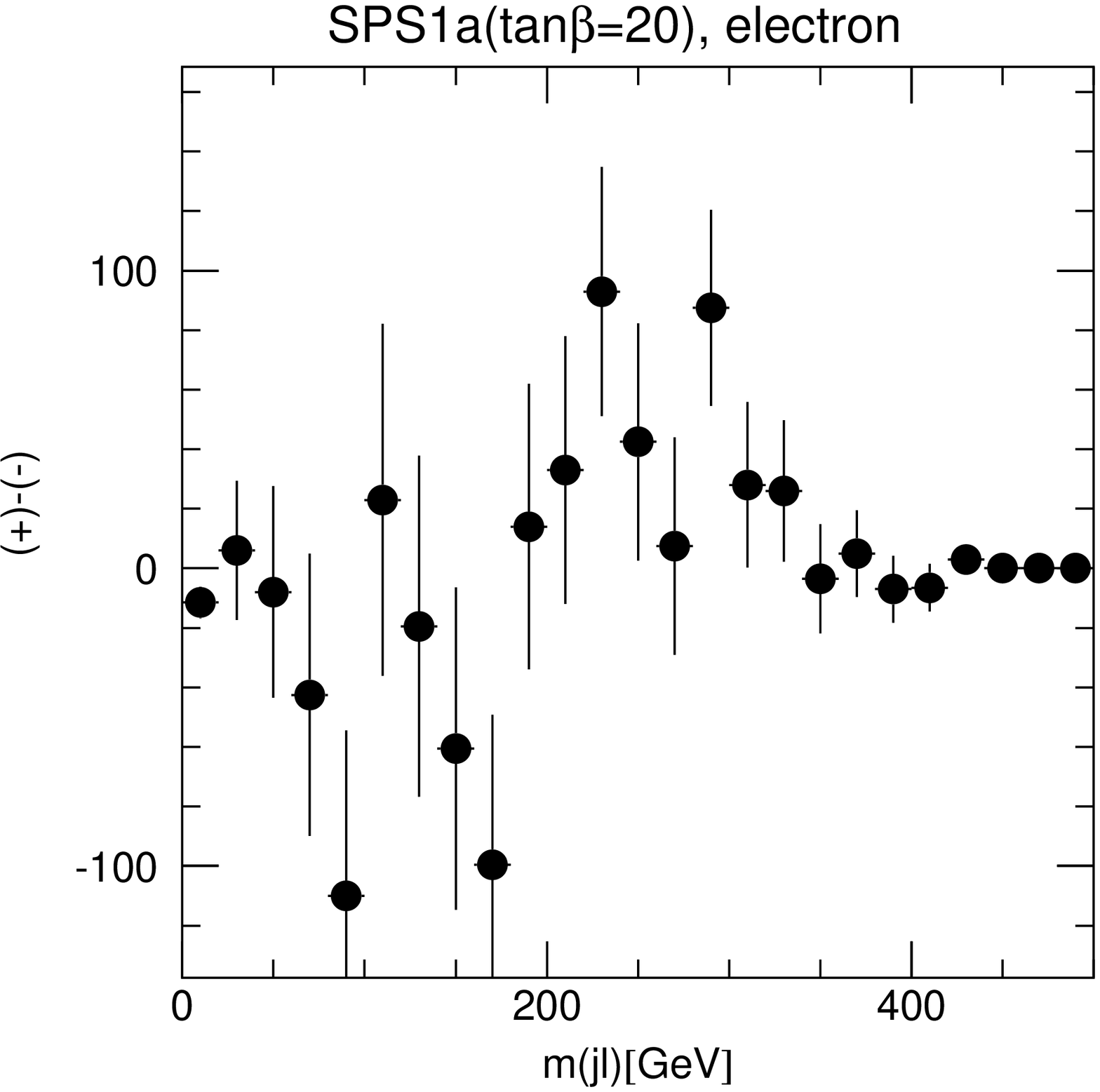}
\includegraphics[height=5cm,clip]{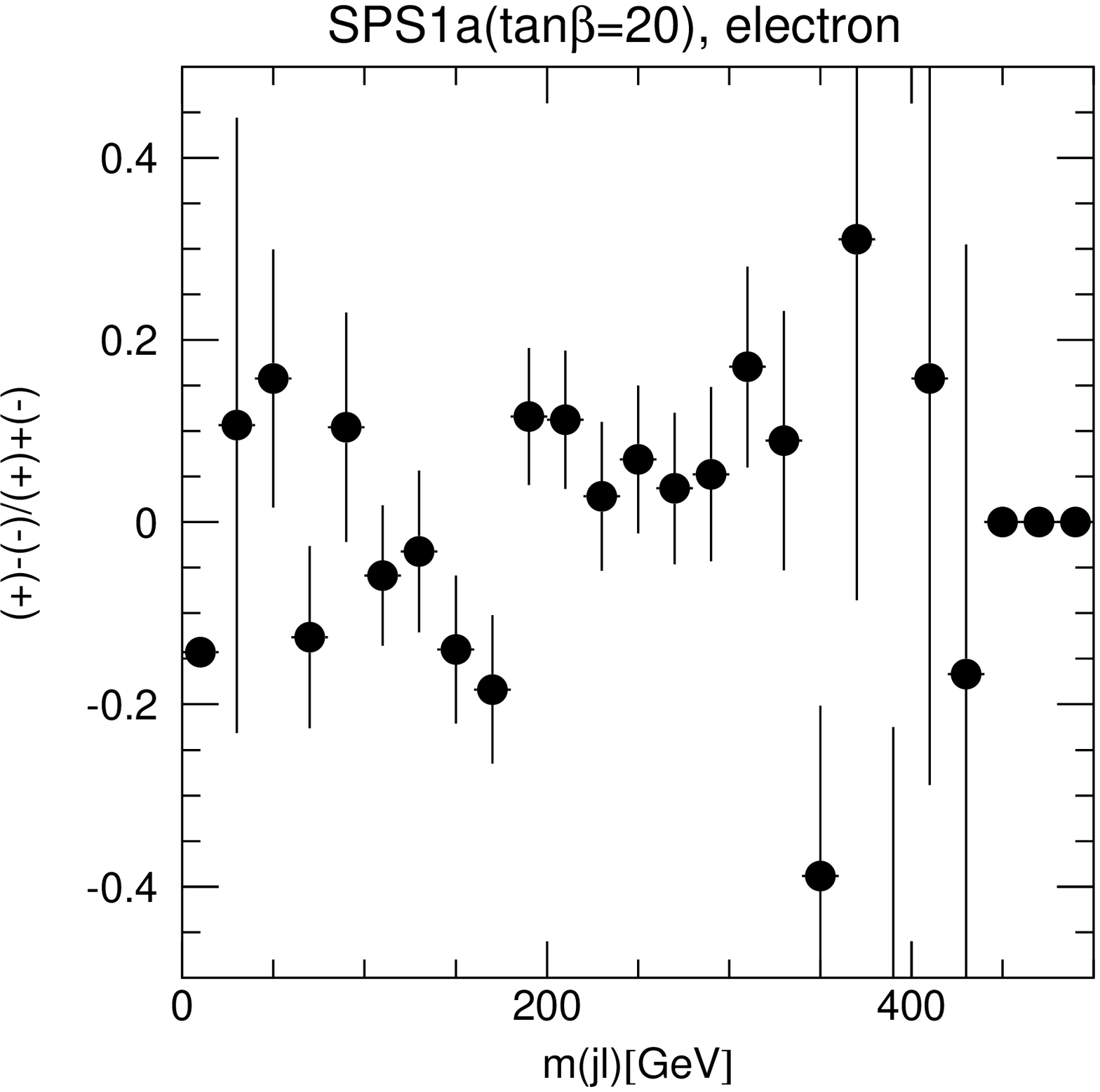}
\includegraphics[height=4.5cm,clip]{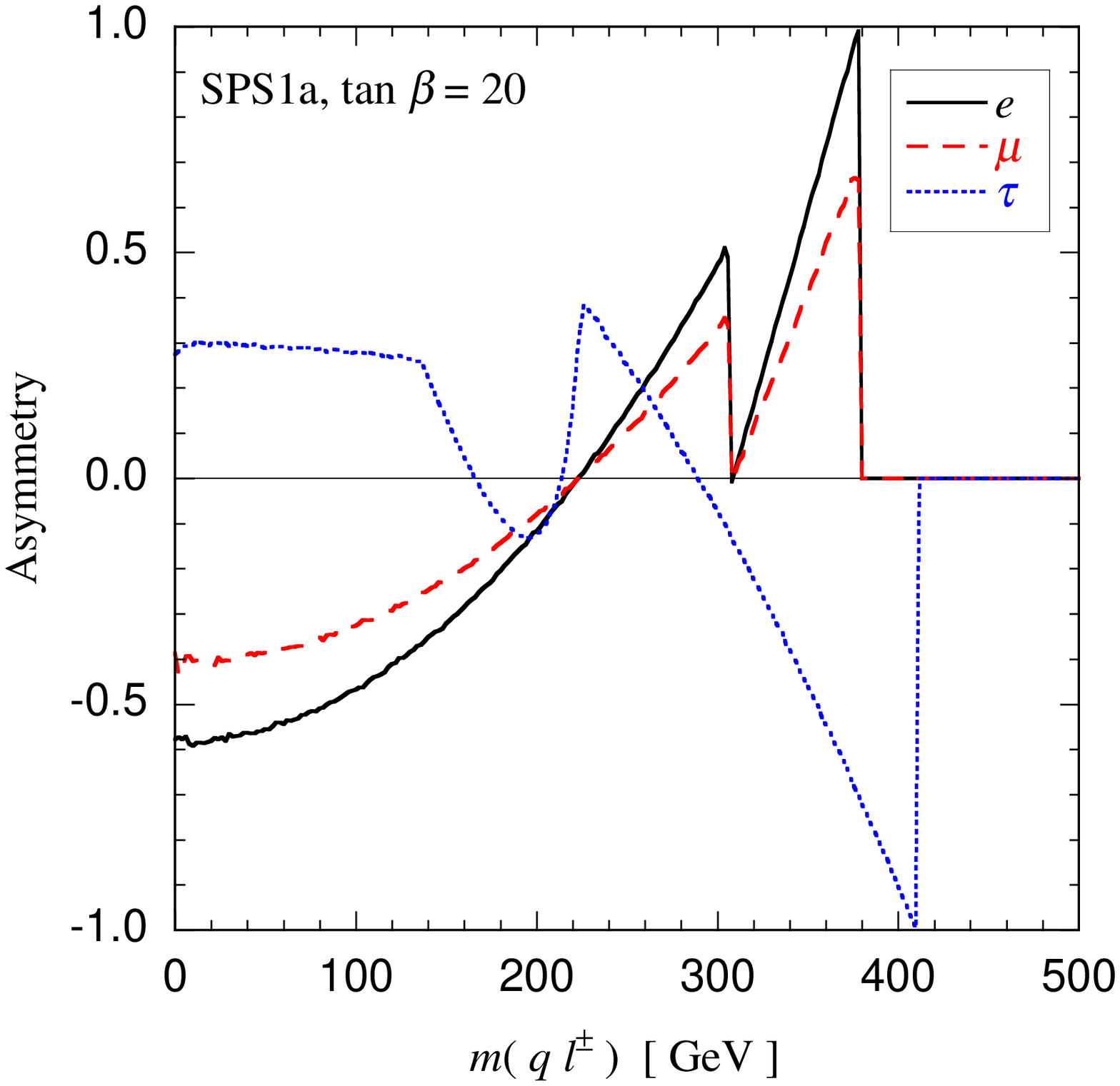}
\\
a)\hskip 5cm b) \hskip 5cm c)
\end{center}
\caption{
a) The reconstructed $N_{\rm sig}(jl^+)-N_{\rm sig}(jl^-)$ as the function of $m(jl)$,
and b) the charge asymmetry for $\int dt {\cal L} =300 ~{\rm fb}^{-1}$
at SPS1a($\tan\beta=20$). 
c) The charge asymmetry in the  $m(ql)$ distribution at
SPS1a($\tan\beta=20$).
}
\label{largetanb}
\end{figure}

\subsection{The non-universal effects}
In this subsection, we  investigate the statistical significance of the non-universal
effect between $e$ and $\mu$ for large $\tan\beta$.
The deviation may be detectable with enough statistics because 
the systematics associated with the uncertainty of $\tilde{q}$ 
production and decay distributions are expected to cancel when 
one takes the $\mu/e$ ratio of the distributions.
On the other side, the measurement will be extremely sensitive to 
systematic uncertainties on the relative efficiency and 
acceptance for electrons and muons. 

The region of the parameters where we can expect large deviation is
rather limited if we assume universal mass spectrum.
This is because $\tilde{\tau}_1$ is lighter than the other sleptons and
it is more $L$-$R$ mixed.
$\tilde{\chi}^0_2$ decays into $\tilde{\tau}_1$ dominantly for large
$\tan\beta$, and $Br(\tilde{\chi}^0_2\to l\tilde{l}_R)$ ($l=e, \mu$) is
suppressed.
As an example, $Br(\tilde{\chi}^0_2\to \tilde{e}e)$, the number of
accepted signal events and the total opposite sign two lepton events
including backgrounds are listed for SPS1a ($\tan\beta=10,20$) in Table
\ref{sigbg} for $60~{\rm fb}^{-1}$.
Number of signal for $\tan\beta=20$ is reduced by a factor of 5, while the
number of backgrounds only reduced by a factor of 2.
The backgrounds dominantly come from $\tau$ leptons from 
Table \ref{sigbg}, we can see that the statistical error of $e^+e^-$ 
ratio $\Delta N({\rm sig})/N({\rm sig})\sim$
$ \sqrt{N_{\rm OS}}/N({\rm sig})\sim 0.02$ for 300 fb$^{-1}$. On the other hand,  the
statistical significance of the charge asymmetry is  3.3 at 
SPS1a($\tan\beta=20$). See Table~ 
\ref{number}.

\begin{table}
\begin{tabular}{|c||c|c|c|c|}
\hline
point & $Br(\tilde{\chi}^0_2\to e\tilde{e})$
 & $N_{\rm sig}$($e$ and $\mu$)  & $N_{\rm OS}$& 
\ \ $\left(\frac{Br(\mu)}{Br(e)}-1\right)
\left(\frac{\Delta N({\rm sig})}{N({\rm sig})}\right)^{-1}$\ \ 
\\
\hline
$\tan\beta=10$ & 6.3\% & $1.39 \times 10^4$ & \ $2.68 \times 10^4$\ & 5.4 \\
\hline
$\tan\beta=20$ & 1.2\%& $0.28 \times 10^4$ & \ $1.02\times 10^4$\ &8.5   \\
\hline
\end{tabular}
\caption{
The $Br(\tilde{\chi}^0_2\to e\tilde{e})$ and accepted number of events
for SPS1a $\tan\beta=10,20$.
$N_{\rm sig}$ is the number of $l^+l^-$ events after $e^{\pm}\mu^{\mp}$
subtraction, while $N_{\rm OS}$ is the total odd sign two lepton events.
The expected deviation of $Br(\tilde{\chi}^0_2\rightarrow \tilde{\mu}\mu)$
from that for $\tilde{e}e$ is compared with the statistical error of the number of  the 
$e^+e^-$ events for $\int dt{\cal L}=300$ fb$^{-1}$.}
\label{sigbg}
\end{table} 
 
When
$\tilde{\mu}$ left-right mixing is taken into account,
$Br(\mu)/Br(e)=1.17$ and $A(\mu)/A(e)=0.70$ for $\tan\beta=20$ (Table~
\ref{tab:tanbeta}).
The statistical significance of the deviation of the branching ratio
$[Br(\mu)/Br(e)-1]/$$[\Delta N({\rm sig})/N({\rm sig})]$
is around 8.5 for 300 fb$^{-1}$. The significance of the deviation of 
charge asymmetry $S\times \vert(A(\mu)-A(e))/A(e)\vert$ 
is only around 1 for 300 fb$^{-1}$. 

The statistical significance at the LHC may be compared with that at the
future linear collider (LC) at $\sqrt{s}=500$~GeV, where
$\tilde{\chi}^0_2\tilde{\chi}^0_2$ cross sections is
$\sim 200~{\rm fb}^{-1}$ for a left handed beam \cite{Desch:2003vw}.
There are 400 $\tilde{\chi}^0_2$$\tilde{\chi}^0_2$ events followed by
$\tilde{\chi}^0_2 \to e\tilde{e}$ for $100~{\rm fb}^{-1}$ at
SPS1a($\tan\beta=20$).
Assuming no background, the deviation of branching ratios around 5\% may
be detectable.

Finally we discuss  the decay distribution of 
$\tilde{\chi}^0_2 \to \tau\tilde{\tau} \to \tau\tau\tilde{\chi}^0_1$.
Ideally the signal distribution
would show charge asymmetry near the end point 
as can be seen in Fig. \ref{asymmetry} c). 
However, as a $\tau$ lepton decays into a meson and $\nu_{\tau}$ or
$l\nu_{l}\nu_{\tau}$, one cannot measure the $\tau$ lepton
momentum directly.
When $\tau$ decays into leptons, the arising charged lepton is soft, and
they typically have less than 1/3 of the $\tau$ lepton energy.
Such large smearing of the energy is not suitable for the charge
asymmetry study.
On the other hand, $\tau$ decay into a meson is a two body decay, and
the meson might be detected as a low multiplicity isolated high $p_T$
jet.
A QCD jet might also be mis-identified as a $\tau$ jet, and the
probability is the function of $\tau$ tagging efficiency
$\epsilon_{\tau}$ and $\tau$ jet $p_T$.
QCD events become large background, however they can be  removed by
subtracting the sum of the distribution of $m(j_{\tau^-} j_{\tau^-})$ and
$m(j_{\tau^+} j_{\tau^+})$ from that of $m(j_{\tau^+} j_{\tau^-})$.

The identification of the hadronic $\tau$ decays has a strong
dependence on the details of the detector performance. 
In the detector simulator ATLFAST, the detection efficiency of the
$\tau$ jet and the rate of fake $\tau$ jet is estimated by using
parameterizations of the full simulation data, which are described in
Ref.~\cite{Hinchliffe:2000np}.
However, the misidentification probability of the 
charge of the $\tau$ jet and the charge distribution of 
fake $\tau$ jets have not been fully accounted for
in the parameterized simulation.
We 
therefore do not study the charge asymmetry in $\tau$  jets 
in this paper.

\section{Discussion}
\label{sec:discussion}

In this paper, we study if SUSY study at the LHC can provide an
information on the chiral nature of the sfermions.
We especially study the charge asymmetry of $m(jl^+)$ and $m(jl^-)$
distribution which comes from the the decay cascade $\tilde{q}\to
\tilde{\chi}^0_2$ $\to \tilde{l}\to \tilde{\chi}^0_1$, where $\tilde{l}$
can be either $\tilde{l}_R$ or $\tilde{l}_L$.

When $\tilde{\chi}^0_2$ is wino like, the asymmetry $N(jl^+)-N(jl^-)$
is negative near the $m(jl)$ end point for pure $\tilde{l}_L$, while it
is positive for pure $\tilde{l}_R$.
We find that the charge asymmetry is significant for two representative
mSUGRA points, SPS1a and SPS3, where slepton masses are lighter than
$m_{\tilde{\chi}^0_2}$.

For $\tilde{\chi}^0_2\to \mu\tilde{\mu}_1$, the branching ratio and
$jl$ distributions may depend on the relation among the left and right
couplings of $\tilde{\chi}^0_2 \tilde{\mu}\mu$ interactions.
The branching ratio and charge asymmetry are proportional to
$|L_{21}^{\mu}|^2 + |R_{21}^{\mu}|^2$ and
$|L_{21}^{\mu}|^2 - |R_{21}^{\mu}|^2$, respectively.
When $\tilde{\chi}^0_2$ is wino like, $L_{21}^{\mu}$ is
proportional to small left-right mixing of $\tilde{\mu}$, and
$\theta_{\mu}$, while $R_{21}^{\mu}$ is proportional to the small
bino component $(U_N)_{21}$ of $\tilde{\chi}^0_2$.
For large $\tan\beta$,
$\cos\theta_{\mu}$ becomes comparable to $(U_N)_{21}$ in mSUGRA points,
leading to the non-universalities in $\Gamma(l)$ and $A(l)$.
We have shown that the LHC can detect $e$-$\mu$ non-universality at
SPS1a($\tan\beta=10$-$20$) for $\int dt {\cal L}=300$~fb$^{-1}$ 
if detection efficiencies of $e$ and $\mu$ 
are understood at the LHC.

The discovery of supersymmetry and determination of the low energy
Lagrangian are the goal of SUSY study at future colliders.
So far, many physics studies have focused on the determination of SUSY
parameters at benchmark points.
On the other hand, the charge asymmetry in the $jl$ distribution provides
an information on the chiral structure of the MSSM, which is constrained
by nature of supersymmetry.
Especially $\tilde{\mu}$ and $\tilde{e}$ non-universality is a direct
evidence of $\tilde{\mu}_L$-$\tilde{\mu}_R$ mixing due to the $F$ term.

\acknowledgments{
We thank the ATLAS collaboration members for useful discussions.
We have made use of the physics analysis framework and tools which are
the result of collaboration-wide efforts.
We especially thank Dr. Polesello, Dr. Barr and Dr. Tovey 
for careful reading 
of the manuscript.
This work is supported in part by the Grant-in-Aid for Science Research,
Ministry of Education, Culture, Sports, Science and Technology (MEXT) 
of Japan (No.~11207101 and
No.~15340076 for K.~K.\ and No.~14540260 and No.~14046210 for M.~M.~N.).
The works of T.~G.\ and M.~M.~N.\ are supported in part by the
Grant-in-Aid for the 21st Century COE ``Center for Diversity and
Universality in Physics'' 
form the MEXT of Japan.
}

\end{document}